\RequirePackage{fix-cm}
\documentclass[smallextended]{svjour3}       
\smartqed  
\usepackage{graphicx}

\usepackage{amsmath}
\usepackage{amssymb} 

\usepackage{listings}
\usepackage{xcolor}

\definecolor{codegreen}{rgb}{0,0.6,0}
\definecolor{codegray}{rgb}{0.5,0.5,0.5}
\definecolor{codepurple}{rgb}{0.58,0,0.82}
\definecolor{backcolour}{rgb}{0.95,0.95,0.92}

\lstdefinestyle{mystyle}{
    backgroundcolor=\color{backcolour},   
    commentstyle=\color{codegreen},
    keywordstyle=\color{magenta},
    numberstyle=\tiny\color{codegray},
    stringstyle=\color{codepurple},
    basicstyle=\ttfamily\footnotesize,
    breakatwhitespace=false,         
    breaklines=true,                 
    captionpos=b,                    
    keepspaces=true,                 
    numbers=left,                    
    numbersep=5pt,                  
    showspaces=false,                
    showstringspaces=false,
    showtabs=false,                  
    tabsize=2
}

\usepackage{float}

\usepackage{booktabs}
\usepackage{makecell}

\usepackage{tcolorbox}

\usepackage[british]{babel}

\usepackage{soul}        
\usepackage[colorinlistoftodos,prependcaption,textsize=tiny]{todonotes}

\usepackage{scalerel}
\usepackage{tikz}
\usetikzlibrary{svg.path}

\definecolor{orcidlogocol}{HTML}{A6CE39}
\tikzset{
  orcidlogo/.pic={
    \fill[orcidlogocol] svg{M256,128c0,70.7-57.3,128-128,128C57.3,256,0,198.7,0,128C0,57.3,57.3,0,128,0C198.7,0,256,57.3,256,128z};
    \fill[white] svg{M86.3,186.2H70.9V79.1h15.4v48.4V186.2z}
                 svg{M108.9,79.1h41.6c39.6,0,57,28.3,57,53.6c0,27.5-21.5,53.6-56.8,53.6h-41.8V79.1z M124.3,172.4h24.5c34.9,0,42.9-26.5,42.9-39.7c0-21.5-13.7-39.7-43.7-39.7h-23.7V172.4z}
                 svg{M88.7,56.8c0,5.5-4.5,10.1-10.1,10.1c-5.6,0-10.1-4.6-10.1-10.1c0-5.6,4.5-10.1,10.1-10.1C84.2,46.7,88.7,51.3,88.7,56.8z};
  }
}

\newcommand\orcidicon[1]{\href{https://orcid.org/#1}{\mbox{\scalerel*{
\begin{tikzpicture}[yscale=-1,transform shape]
\pic{orcidlogo};
\end{tikzpicture}
}{|}}}}
\usepackage{hyperref}

\usepackage{orcidlink}

\usepackage[htt]{hyphenat}

\usepackage{xurl}

\begin{document}

\title{
Beyond the Tip of the Iceberg: Understanding SATD in Dockerfiles through the Lens of Co-evolution
}

\author{
Wei~Minn~\orcidlink{0000-0002-3191-9795} \and
Yan~Naing~Tun~\orcidlink{0009-0009-2899-4637} \and
Biniam~Demissie~\orcidlink{0000-0002-5369-5235} \and
Rui'ang~Hu\and \\
Jiakun~Liu~\orcidlink{0000-0002-7273-6709}\footnote{Corresponding author.} \and  
Mariano~Ceccato~\orcidlink{0000-0001-7325-0316} \and
Lwin~Khin~Shar~\orcidlink{0000-0001-5130-0407} \and
David~Lo~\orcidlink{0000-0002-4367-7201}
}

\institute{
Wei Minn, Yan Naing Tun, Lwin Khin Shar, and David Lo \at
Singapore Management University\\
\email{wei.minn.2023@phdcs.smu.edu.sg, yannaingtun@smu.edu.sg, lkshar@smu.edu.sg, davidlo@smu.edu.sg}  \\
\and
Biniam Fesseha Demissie \at
Technology Innovation Institute, UAE \\
\email{biniam.demissie@tii.ae} \\
\and
Rui'ang Hu, Jiakun Liu \at
Harbin Institute of Technology, China \\
\email{ruianghu@foxmail.com, jiakunliu@hit.edu.cn} \\
\and
Mariano Ceccato \at
University of Verona, Italy \\
\email{mariano.ceccato@univr.it} \\
}

\date{Received: date / Accepted: date}

\maketitle

\begin{abstract}

Dockerfiles enable the creation of portable container-based execution environments for the application code, and have become an important part of the modern software development process. 
As Dockerfiles are a form of Infrastructure-as-Code (IaC), they can include temporary workarounds and other suboptimal implementations, leading to the accrual of technical debt that affects their reliability, security, and maintainability in the future.
Prior work characterized technical debt in Dockerfiles by analyzing self-admitted technical debt (SATD) in Dockerfile comments and the surrounding file chunks.
This single-file view is incomplete 
since source code evolution involves changes across different types of software artifacts such as production, test, build, and other configuration files. 
Thus, we address this gap by studying SATD events in Dockerfiles alongside the related source code.

We construct a dataset of $1{,}316$ SATD instances from $393$ Docker Hub--GitHub linked repositories with full Dockerfile histories, manually validate and categorize the SATD instances using the taxonomy of Azuma et al., and annotate whether each admission and repayment event is coupled to non-Dockerfile co-changes.
Our findings complement the existing Dockerfile SATD literature's singular dimension of SATD subtypes with source code-side co-evolutionary dimension for extra contextual insights.
More specifically, we find that
approximately $27\%$ of admission events and $40\%$ of repayment events are coupled to non-Dockerfile artifacts, and coupling sources are subtype-specific.
We also observed that coupled SATD in general are repaid significantly faster overall ($p=0.0201$), while coupled SATD regarding missing functionalities persists longer than its isolated counterparts; 
Lastly, we conducted open and axial coding of coupled SATD events, and we observe that
external dependency issues, more particularly regarding unreleased upstream packages and bug fixes, are the most common cause of admission triggers in the source code;
we also observe that architectural refactoring is the most common prerequisite for the repayment of SATD in Dockerfiles.
These findings indicate that both practitioners (e.g. developers and project managers) and SATD researchers should integrate the source code-side co-evolution, rather than the single-file view, as the primary unit of analysis. For example, automated SATD detection and repair tools should integrate cross-artifact signals, and project-level remediation triage and task delegation should follow the coupling source rather than the file location.

\end{abstract}

\vspace{-1.25em}
\section{Introduction}
\label{sec:introduction}

Dockerfiles have become an integral part of modern software development as they allow software systems to be built, packaged, and executed in portable containerized environments.
They enable convenient dependency installation and environment configuration, consistent builds, and deployment portability across production settings~\cite{noauthor_docker_nodate}.
As such, Dockerfiles serve as the crucial interface between application code and its development environment, making them a widely-adopted technology in the Software Development Life Cycle (SDLC)~\cite{noauthor_what_2024}.

Various works have studied Dockerfiles from perspectives akin to the study of conventional source code analyses.
For example, existing works have mined recurring specification patterns and rules from large Dockerfile corpora to characterize common practices and infer usage regularities~\cite{henkel_learning_2020,zhou_drive_2023}.
Other studies have focused on Dockerfile refactoring, proposing techniques to identify and restructure suboptimal constructs to improve maintainability~\cite{ksontini_drminer_2024}.
A substantial body of work has examined Dockerfile smells, including their characterization, detection, repair, and empirical consequences~\cite{rosa_fixing_2024,bui_dockercleaner_2023,durieux_parfum_2023,rosa_not_2024,durieux_empirical_2024,wu_characterizing_2020}.
More recently, research has explored automated Dockerfile repair, aiming to synthesize fixes for faulty or undesirable specifications~\cite{shabani_dockerfile_2025}.

\begin{figure}[H]
    \centering
    \includegraphics[width=.9\textwidth, trim={0.6cm 1cm 0.6cm 1cm}, clip]{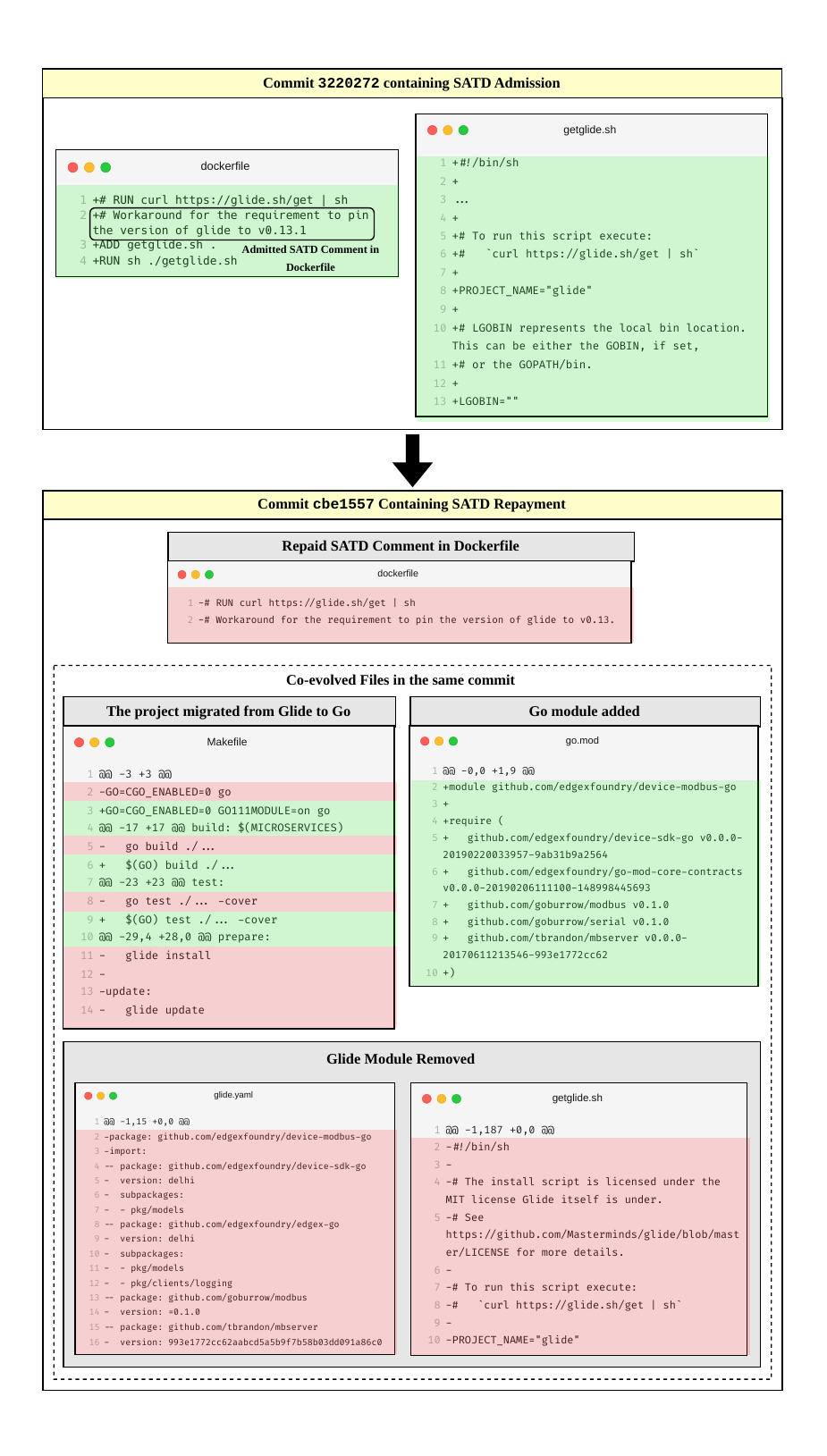}
    \caption{In commit \texttt{3220272} of \texttt{edgexfoundry/device-modbus-go} project, an SATD comment (about using Glide package manager) was added to the Dockerfile, and, in the same commit, \texttt{getglide.sh} script for configuring Glide is added. Later in commit \texttt{cbe1557}, the Dockerfile shows the removal of the SATD comment; the four co-changed files (\texttt{Makefile}, \texttt{go.mod}, \texttt{getglide.sh}, and \texttt{glide.yaml}) in the same commit are semantically linked to the repayment.}
    \label{fig:motivation}
\end{figure}

Among these lines of work, self-admitted technical debt (SATD) in Dockerfiles has received only limited attention.
Within our knowledge, the only dedicated study is the empirical investigation by Azuma et al., which identifies and categorizes SATD comments appearing in Dockerfiles~\cite{azuma_empirical_2022}.
While it offers valuable insights into the taxonomy of Dockerfile SATD, it analyzes Dockerfile SATD localized only within a single file.
Prior work on IaC-source co-evolution has long established that IaC artifacts and application code are closely interconnected and evolve together~\cite{jiang_co-evolution_2015,wu_dockerfile_2020,mcintosh_mining_2014}.
Despite this understanding of IaC-code coupling, no existing work has investigated whether SATD is admitted, maintained, or repaid in connection with changes outside the SATD-occurring artifact.
For instance, the latest work on automated SATD repayment~\cite{mastropaolo_towards_2024} only uses the code chunks containing SATD comments to fine-tune their model.
Their model is only able to resolve $2.30\%$ of the cases on first attempt and $8.10\%$ of the cases in $10$ attempts, 
thus demonstrating that there is still a long way to go in the topic of automated SATD repayment.

Therefore, we hypothesize that SATD comments, more specifically in Dockerfiles, are not merely local workarounds. Rather, they are the resultant of limitations, constraints, and workarounds in related source-code side artifacts such as application logic code, test code, dependency manifests, and other configuration files.
For example, there may be missing functionality inside logic code, compatibility issues regarding dependency version changes, and constraints in the CI/CD pipeline and infrastructure.
Thus, the repayment of such kind of Dockerfile SATD is contingent upon source-code side improvement in terms of implementing new application logic, upgrading dependencies, and updating configuration values.

Figure~\ref{fig:motivation} shows an SATD comment admitted in a Dockerfile at commit \texttt{3220272} in the \texttt{edgexfo}-\texttt{undry/device-modbus-go} project on GitHub\footnote{https://github.com/edgexfoundry/device-modbus-go/commit/3220272}.
The SATD comment and the surrounding code chunk implies a version-related workaround caused by a Glide-related script somewhere in the project.
The commit-level inspection reveals the addition of the \texttt{getglide.sh} script which was mentioned by the SATD comment in the Dockerfile.
Thus, the broader project-level build system-related workaround is reflected in this Dockerfile SATD event. 

Figure~\ref{fig:motivation} also shows the removal (repayment) of the SATD comment in the same Dockerfile in a follow-up commit \texttt{cbe1557}\footnote{https://github.com/edgexfoundry/device-modbus-go/commit/cbe1557}.
In this commit, Glide-related artifacts (\texttt{getglide.sh}, and \texttt{glide.yaml}) are totally removed, while Go-related configuration file (\texttt{go.mod}) is added.
Moreover, in the \texttt{Makefile}, Glide-related statements make way for the Go-related ones.
Thus, the repayment of the Dockerfile SATD comment happens only when the project migrated from Glide to Go package management system.

Motivated by such correlated changes between Dockerfile SATD and broader project artifacts, we constructed a large-scale dataset of Docker Hub-GitHub linked projects that contain Dockerfiles, and manually annotated SATD lifecycle events and coupling relationships.
Our study is guided by the following research questions:

\begin{itemize}
    \item \textbf{RQ1: How often are different SATD subtypes coupled to non-Dockerfile artifacts during admission and repayment? What coupling sources dominate per subtype?}
    This RQ adds an additional dimension to the existing empirical landscape as existing literature characterizes only the \emph{content} of Dockerfile SATD comments, and not whether those SATD events are primarily isolated to the Dockerfile or linked to changes elsewhere in the repository.
    27\% of admission events and 40\% of repayment events are coupled with non-Dockerfile artifacts, which suggests that the Dockerfile-only view is limited when studying automated SATD annotation and repayment.
    Coupling sources are subtype-specific: code-centric subtypes co-evolve primarily with application logic files, test-related subtypes with CI/CD artifacts, and design or size-reduction subtypes with infrastructure files.

    \item \textbf{RQ2: Do coupled SATD instances differ from isolated ones in time-to-repayment?}
    In order to study SATD repayment dynamics, we track SATD instances over time, identifying admission, persistence, and repayment events, and compare the survival behavior of coupled versus isolated debt using Kaplan-Meier analysis and log-rank tests~\cite{kaplan_nonparametric_1958,peto_asymptotically_1972}.
    Counter-intuitively, coupled SATD instances are repaid significantly faster overall ($p=0.0201$), with \textit{Defect/Workaround} showing the strongest effect ($p=0.0084$), suggesting that developers assign higher priority to cross-artifact debt.
    By contrast, \textit{Code/MissingFunctionality} coupled instances tend to persist longer, indicating that missing-feature debt requiring upstream effort is harder to resolve even once the coupling is recognized.

    \item \textbf{RQ3: What cross-artifact changes trigger Dockerfile SATD admissions and enable their repayments?}
    In order to understand the full lifecycle of Dockerfile SATD, we need to supplement with the analysis of co-changed artefacts at both admission and repayment time.
    We taxonomized coupled admission and repayment events using open and axial coding~\cite{pearson_vii_1895} to derive different types of admission triggers and repayment requisites.
    For admissions, we identify four themes: \textbf{External Dependency Constraints}, \textbf{Compatibility and Environment Issues}, \textbf{Incomplete Implementation}, and \textbf{Maintenance Overhead}.
    For repayments, we identify three requisite themes: \textbf{Upstream Progression}, \textbf{Internal Refactoring}, and \textbf{Feature Completion}.
\end{itemize}

\noindent This paper makes the following contributions:
\begin{itemize}
    \item \textbf{(New paradigm)} We introduce a \emph{co-evolutionary dimension} to the study of Dockerfile SATD, extending prior single-artifact views with the perspective of different types of co-evolved source files.
    \item \textbf{(Dataset)} We build and curate a large-scale dataset of Docker Hub-GitHub linked repositories with full Dockerfile histories, SATD candidate extraction, manually labeled SATD instances along with their SATD subtypes, and 
    manually labeled semantic coupling candidates~\cite{minn_dockerfile_nodate}.
    \item \textbf{(Coupling and Survival Analysis)} We analyze coupling sources, and repayment dynamics of Dockerfile SATD and show that coupled and isolated debt instances differ systematically in their time-to-repayment behavior depending on their SATD subtypes.
    \item \textbf{(Taxonomy)} We present a qualitative taxonomy covering cross-artifact \emph{admission triggers} and \emph{repayment requisites}, explaining why Dockerfile SATD is introduced and what enables its removal.
\end{itemize}

\begin{figure}[t] 
    \centering
    \includegraphics[width=.9\textwidth, trim={0.65cm 0.25cm 0.65cm 0.25cm}, clip]{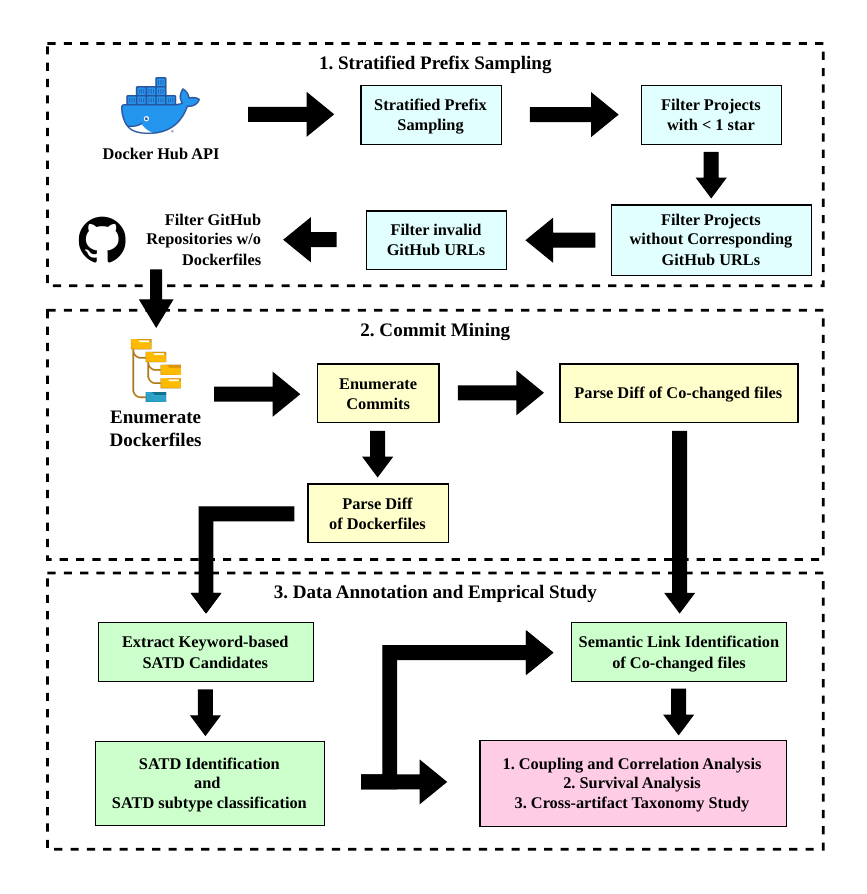} 
    \caption{Overview of our dataset construction pipeline: first, we perform stratified prefix sampling to scrap all the Docker Hub projects with at least 1 start, and extract URLs to valid corresponding GitHub repositories. Then, for each Dockerfile in the GitHub repositories, we look into all the commits that touched the Dockerfile, and extracted diff snippets of the Dockerfile and co-changed files. Then, we perform keyword-based SATD candidate comment extraction from the Dockerfiles. These candidates are then verified by human annotators as part of SATD identification and subtype classification step. The annotators also label semantic link between the SATD in Dockerfile and the co-changed file. Finally, we use the labelled SATD and semantic link data for our empirical analyses.}
    \label{fig:overview}
\end{figure}

\section{Dataset Construction}
\label{sec:dataset}

Understanding Dockerfile SATD through a co-evolutionary lens requires a dataset that captures not only the SATD comments themselves, but also the surrounding co-evolved source files.
We separate (i)~\emph{SATD identification and subtype classification} for Dockerfile comments from (ii)~\emph{event-level coupling assessment} for SATD admissions and repayments.

\subsection{Stratified Prefix Sampling}
\label{subsec:prefix-sampling}

In order to explore a representative set of Docker-related projects, we query Docker Hub repositories using string prefixes where we query the Docker Hub API~\cite{noauthor_docker_nodate-1} to retrieve all Docker Hub projects that start with all possible combinations of 3-character prefixes.
Table~\ref{tab:dataset-flow} summarizes the number of retained project throughout the dataset construction pipeline.
Then, we obtain $151{,}295$ Docker Hub projects with at least one star. Thereby, we exclude $14$ million $0$-star Docker Hub projects, making the scope of our project feasible. 
For each retained Docker Hub projects, we extract GitHub URLs from the project descriptions and metadata using a documented set of regular expressions.
There are $4{,}300$ Docker Hub projects that has metadata that contain links to their corresponding GitHub repositories.
Extracted URLs are normalized to canonical Git Hub repository identifiers of the form \texttt{<owner>/<repo>}. 
We discover $3{,}021$ valid corresponding GitHub repositories, and excluded the Docker Hub repositories that cannot be linked to valid public GitHub repositories from subsequent mining.
$2{,}962$ out of the $3{,}021$ corresponding GitHub repositories contain Dockerfiles. and $393$ out of the $2{,}962$ Dockerfile-containing GitHub repositories contain SATD candidate comments.

\begin{table}[t]
\centering
\caption{Dataset construction statistics. }
\label{tab:dataset-flow}
\begin{tabular}{clr}
\toprule
\textbf{Stage} & \textbf{Description} & \textbf{Count} \\
\midrule
1 & Docker Hub projects after star $\geq 1$ filter & $151,295$ \\
2 & Docker Hub projects with extractable GitHub URL & $4,300$ \\
3 & Corresponding GitHub repositories & $3,021$ \\
4 & Corresponding GitHub repositories with Dockerfile history & $2,962$ \\
5 & Corresponding GitHub repositories with SATD candidate comments & $393$ \\
\bottomrule
\end{tabular}
\end{table}

\subsection{Heuristics for Artifact-Type Classification}
\label{subsec:artifact-type}

In order to study the coupling source of the different types of SATD, we need to classify the different types of co-changed files. 
Following McIntosh et al.~\cite{mcintosh_empirical_2011} and Jiang \& Adams~\cite{jiang_co-evolution_2015} that studied co-evolution between IaC and source code, we leverage known file naming patterns to classify the source file type into \textsc{Logic}, \textsc{Test}, \textsc{Build/Dependency}, \textsc{CI/CD}, and \textsc{Infrastructure}:
\begin{itemize}
    \item \textsc{Test} $\leftarrow$ file path contains \texttt{test} or \texttt{spec}.
    \item \textsc{Logic} $\leftarrow$ file path ends in common source-code extensions (e.g., \texttt{.java}, \texttt{.py}, \texttt{.go}, \texttt{.js}).
    \item \textsc{Build/Dependency} $\leftarrow$ file path matches well-known build and dependency manifests such as \texttt{pom.xml}, \texttt{package.json}, \texttt{requirements.txt}, \texttt{go.mod}, \texttt{build.gradle}, \texttt{Cargo.toml}, or \texttt{Makefile}.
    \item \textsc{Infrastructure} $\leftarrow$ file path with extension (e.g., \texttt{.tf}, \texttt{.tfvars} for Terraform; \texttt{.yaml}/\texttt{.yml} files matching Kubernetes, Ansible, or CloudFormation naming conventions; \texttt{.toml}, \texttt{.jsonnet}, \texttt{Vagrantfile}) or by path fragment (e.g., \texttt{docker-compose}, \texttt{Dockerfile}).
\end{itemize}
This coarse artifact-type annotation is used in later steps to characterize the dominant coupling sources for each SATD subtype in Section~\ref{sec:results-rq1} (RQ1) and to support the qualitative coding in Section~\ref{sec:results-rq3} (RQ3).

\subsection{Data Annotation}
\label{subsec:data-annotation}

Our manual annotation process involves three labeling tasks: (1)~\textbf{SATD identification} which labels whether a candidate Dockerfile comment is a genuine SATD,
(2)~\textbf{SATD subtype classification} which classifies the SATD instances to a subtype from the Dockerfile SATD taxonomy presented by Azuma et al.~\cite{azuma_empirical_2022}, and 
(3)~\textbf{SATD event coupling classification} which classifies whether a SATD admission or repayment event is semantically linked to co-changed non-Dockerfile files from the same commit.
Table~\ref{tab:annotation-schema} summarizes the annotation tasks, their units of analysis, label sets, and the data shown to annotators.

\paragraph{Annotators and annotation protocol.}
Two annotators, with $12$ and $15$ respective years of programming experience, independently labeled all three tasks.
The annotation for each of the three tasks happens in three phases: (1) \emph{pilot phase}, (2) \emph{reliability phase}, and (3) \emph{production phase}. 
In the \emph{pilot phase}, both annotators independently labeled a shared sample of $60$ instances per task, and discussed disagreements to refine the codebook. 
In the \emph{reliability phase}, both annotators independently labeled a new shared sample of $100$ instances per task, and measured inter-rater agreement statistics in terms of Cohen's $\kappa$~\cite{cohen_coefficient_1960}. 
In the \emph{production phase}, the remaining instances were split evenly between the two annotators, with a final adjudication meeting to resolve conflicting cases.

\begin{table*}[t]
\centering
\caption{Annotation schema used in the study.}
\label{tab:annotation-schema}
\begin{tabular}{p{0.14\linewidth}p{0.17\linewidth}p{0.2\linewidth}p{0.35\linewidth}}
\toprule
Task & Unit of analysis & Labels & Data shown to annotators \\
\midrule
SATD identification & Candidate Dockerfile comment & SATD / NotSATD & Comment text, Dockerfile path, surrounding Dockerfile context, commit message \\
SATD subtype classification & Validated SATD comment & SATD taxonomy subtypes presented by Azuma et al.~\cite{azuma_empirical_2022} & Comment text, Dockerfile path, surrounding Dockerfile context, commit message \\
SATD event coupling classification & SATD admission or repayment event & Coupled / Isolated & SATD comment, Dockerfile context, event type, commit message, co-changed file paths, co-changed file diffs \\
\bottomrule
\end{tabular}
\end{table*}

\subsection{SATD Identification}
\label{subsec:satd-identification}

In order to extract all SATD instances from the GitHub repositories, we enumerate all Dockerfile paths throughout the full commit history for all the repositories. Then, we filter for commits that touch a Dockerfile, and extract comment lines that were added to or removed from the Dockerfiles. 
We treat a comment line inside a Dockerfile as \textit{SATD candidate} if it contains at least one of the phrases from the SATD keyword list such as \texttt{TODO}, \texttt{FIXME}, \texttt{HACK}, \texttt{WORKAROUND}, \texttt{UGLY}, or \texttt{PROBLEMATIC}.
This keyword-based SATD candidate filtering is in line with the prior work in SATD empirical study~\cite{guo_how_2021}, and is specifically validated for Dockerfile domain by Azuma et al.~\cite{azuma_empirical_2022} which sampled 330 Dockerfile comments \emph{without} the SATD keywords and found that only $1.83\%$ of them were SATD. Thus, we consider the use of keyword-based filtering good enough for initial SATD candidate collection.

For each discovered SATD candidate, we further extract the context of the SATD instance: (1) comment text itself, (2) Dockerfile path, (3) surrounding code block preceding and following the comment, (4) commit message, (5) co-changed files in the same commit, and (6) commit hash and timestamp.
Using the information from the extracted SATD context, the annotators independently assign each SATD candidate to one of two labels:
\begin{itemize}
    \item \textbf{SATD}: the candidate explicitly admits a suboptimal, temporary, workaround, or to-be-revisited technical condition as presented by prior works in SATD identification~\cite{guo_how_2021,azuma_empirical_2022}, and
    \item \textbf{Non-SATD}: the candidate does not self-admit technical debt.
\end{itemize}
Following the three-phase protocol described in Section~\ref{subsec:data-annotation}, the two annotators achieved a Cohen's $\kappa$ of $0.66$ on the shared subset in the \textit{reliability phase}, indicating substantial agreement~\cite{landis_measurement_1977}.
As the total number of candidate comments is manageable ($1{,}316$ candidates), we annotate all $1{,}316$ candidates rather than sampling a subset of them. Thus, we avoid potential sampling errors and biases at this SATD identification stage and present a representative comment-level SATD dataset.

\begin{table}[t]
\centering
\caption{SATD Subtypes presented by Azuma et al.~\cite{azuma_empirical_2022} }
\label{tab:satd_subtypes}
\begin{tabular}{ll}
\toprule
SATD Subtype & Description \\
\midrule
Code/Workaround & SATDs on suboptimal implementation\\
Code/MissingFunctionality & SATDs on the lack of functionality inside containers\\
Code/BaseImage & SATDs on the base image\\
Test/IntegrityCheck & SATDs on binary integrity verification\\
Test/ImprovementForTest & SATDs on improvement for testing\\
Defect/Workaround & SATDs on working around bugs in external systems\\
Defect/LatentBug & SATDs on latent bugs\\
Design/SizeReduction & SATDs on size optimization of Docker image\\
Process/Deployment & SATDs on deployment processes\\
Process/Review & SATDs on review of the Dockerfile\\
Unclassifiable & Clear SATD, but not enough information\\
\bottomrule
\end{tabular}
\end{table}

\vspace{-.5em}
\subsection{SATD Subtype Classification}
\label{subsec:satd-classification}

\begin{figure}[htbp]
    \centering
    \includegraphics[width=\textwidth, trim={.45cm 0 .25cm .45cm}, clip]{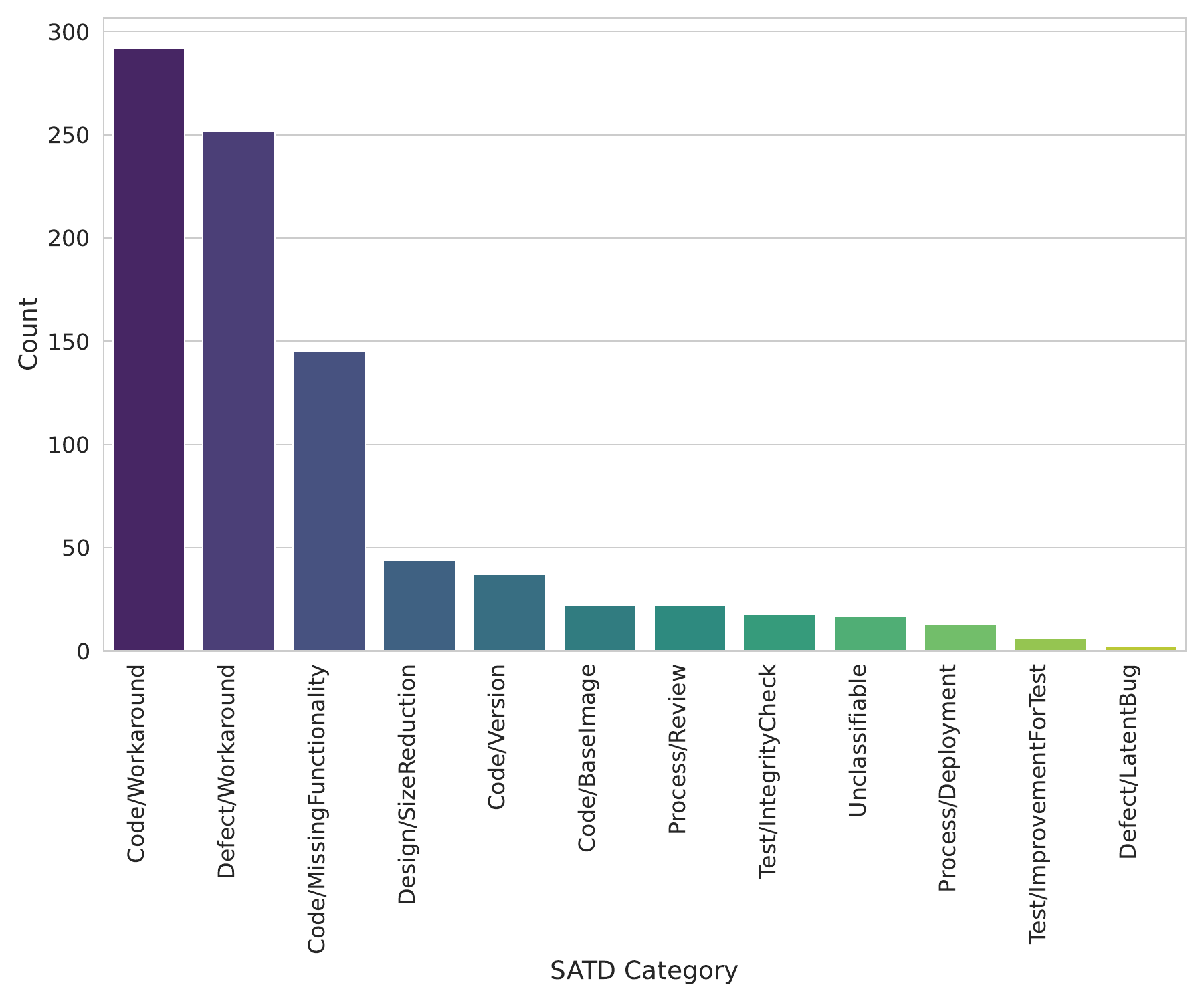}
    \caption{Distribution of manually validated Dockerfile SATD instances across subtypes.}
    \label{fig:prevalence}
\end{figure}

In order to enable analysis of Dockerfile SATD by their subtypes, we classified the labelled \textbf{SATD} instances into Dockerfile SATD subtypes based on the taxonomy presented by Azuma et al.~\cite{azuma_empirical_2022}.
Table~\ref{tab:satd_subtypes} shows the $10$ Dockerfile SATD subtypes from the taxonomy study. 
Following the three-phase protocol described in Section~\ref{subsec:data-annotation}, two annotators classify each SATD instance to exactly one subtype, and discuss disagreements to refine the codebook that comprises the SATD subtype definitions, and their corresponding examples.
In the \textit{reliability phase}, the two annotators achieved a Cohen's $\kappa$ of $0.63$ on the shared subset, indicating substantial agreement~\cite{landis_measurement_1977}.
Figure~\ref{fig:prevalence} summarizes the resulting distribution of validated SATD instances across subtypes, providing an overview of the SATD landscape in our dataset as the context for the coupling and lifecycle analyses in Section~\ref{sec:results-rq1} and~\ref{sec:results-rq2}.

\subsection{SATD Coupling Annotation}
\label{subsec:event-coupling}

To study SATD in Dockerfile from the lens of co-evolution, we need to analyze changes in other co-changed source files in the same commit.
However, we need to determine if the co-changed files are semantically linked (coupled) to the SATD comment inside the Dockerfile or just a coincidental change that is related to a different task in the same commit.
Thus, we need to label which of the SATD events from the identified SATD events in Section~\ref{subsec:satd-identification} are coupled or not.
More precisely, we define coupling as follows:
\begin{quote}
\emph{A SATD event is labeled \textbf{Coupled} if at least one co-changed non-Dockerfile artifact plausibly contributed to the admission of the SATD or to the conditions enabling its repayment. Otherwise, the event is labeled \textbf{Isolated}.}
\end{quote}

For each sampled SATD event, the two annotators refer to the information in the SATD comment text, Dockerfile code chunk context around the SATD, event type (Admission or Repayment), commit message, the list of diff snippets of co-changed non-Dockerfile files in the same commit.
Following the three-phase protocol described in Section~\ref{subsec:data-annotation}, the two annotators independently labeled the shared pilot sample to construct a codebook for semantic linking of co-changed file to the Dockerfile SATD comment. 
Then, in the \textit{reliability phase}, the two annotators achieved a Cohen's $\kappa$ of $0.62$ on the shared reliability subset, indicating substantial agreement~\cite{landis_measurement_1977}. 
Afterwards, in \textit{production phase}, the two annotators split the remaining SATD events, and labeled the coupling statuses of respective part of the dataset.
Lastly, the two annotators held a discussion to manually resolve conflicting labels from \textit{production phase}.

\vspace{-.5em}
\section{Coupling and Correlation Analysis (RQ1)}
\label{sec:results-rq1}

\subsection{Motivation}

As the existing Dockerfile SATD study~\cite{azuma_empirical_2022} characterizes only the content of Dockerfile SATD comments and surrounding code context in the same Dockerfile, 
it overlooks the possibility of Dockerfile SATD being frequently coupled with other artifects in the project such as application code, build system, and other configurations files. 
This would severely limit our understanding of Dockerfile SATD in terms of root causes, and remediation strategies that could only be found inside the source code.
Thus, we first get an overview of the SATD landscape in Dockerfiles in terms of the distribution or prevalence of Dockerfile SATD subtypes.
Then, we examine whether their admission and repayment events occur in isolation or together with semantically linked non-Dockerfile changes.
Finally, we characterize the dominant \emph{coupling sources} for each subtype.

\subsection{Approach}

To explore the prevalence landscape of SATD, we quantified the manually-annotated Dockerfile SATD instances and their subtypes during the dataset construction phase in Section~\ref{subsec:satd-classification}. Figure~\ref{fig:prevalence} shows the distribution of SATD subtypes across all identified instances.

\begin{figure}[htbp] 
    \centering
    \includegraphics[width=\textwidth, trim={.5cm 0 .5cm .25cm}, clip]{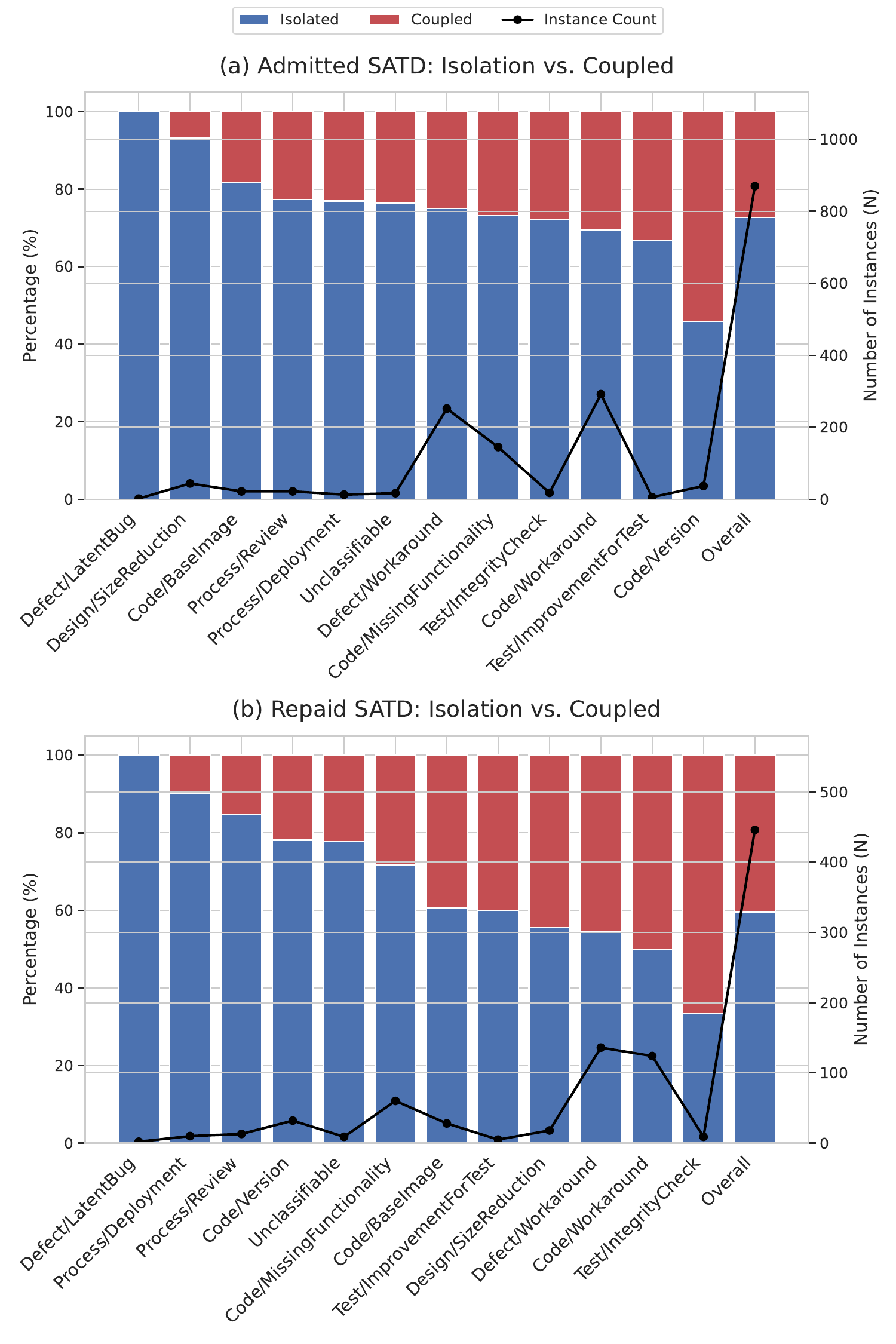} 
    \caption{Isolation versus coupling rates by SATD subtype for (a) admissions and (b) repayments. The line plot indicates the number of instances per subtype.}
    \label{fig:coupling-rates}
\end{figure}

To find out the coupling rates of Dockerfile SATD, we look at two kinds of SATD lifecycle events: \emph{admission} and \emph{repayment}. 
We use the manually-annotated of SATD coupling data from Section~\ref{subsec:event-coupling}, and quantified the proportion of \textit{Coupled} vs. \textit{Isolated} events for each SATD subtype where
\textit{Coupled} denotes events that has at least one semantically-linked file change in the same commit, and \textit{Isolated} if otherwise.
Figure~\ref{fig:coupling-rates} displays the proportion of \textit{Isolated} and \textit{Coupled} events per SATD subtype for both admissions and repayments, together with the corresponding number of instances.

To further investigate the nature of Dockerfile SATD coupling, we analyzed the relationship between SATD subtypes and co-changed source file types using binary correlation analysis.
For every co-changed file in each SATD instance, we mapped the semantically-linked files into five source categories: 
\emph{Build/Dependency}, \emph{CI/CD}, \emph{Infrastructure}, \emph{Logic}, and \emph{Test} using the file path-based heuristics discussed in Section~\ref{subsec:artifact-type}.
For every SATD event, we encoded binary variables for SATD subtype and file types on whether the SATD admission belongs to an SATD subtype, and whether its co-changes at least one file of the file type.
Then we computed the $\phi$ coefficient~\cite{cramer_mathematical_1946} which is a special case of Pearson correlation~\cite{pearson_vii_1895} for binary variables, 
and assessed significance using Fisher's exact test with Benjamini-Hochberg correction~\cite{benjamini_controlling_1995}.
Table~\ref{tab:satd_filetype_correlation} reports SATD subtype-file type correlations ranked by absolute $\phi$, and Figure~\ref{fig:associations} visualizes the $\phi$ coefficients for the five most prevalent SATD subtypes.

\subsection{Results}

\vspace{0.5em}\noindent
\textbf{Dockerfile SATD is dominated by workarounds related to both code and defect themes.} Figure~\ref{fig:prevalence} shows that \textit{Code/Workaround} is the most prevalent subtype (292), followed by \textit{Defect/Workaround} (252) and \textit{Code/MissingFunctionality} (145).
Together, these three categories account for the large majority of all identified SATD instances, far exceeding less frequent categories such as \textit{Design/SizeReduction} (44), \textit{Code/Version} (37), and process- or testing-related debt.
This suggests that the most prevalent forms of Dockerfile SATD are not merely comments about image hygiene or build optimization. 
Rather, they highlight mismatches, workarounds, or limitations in the underlying software system.
This insight is also consistent with the study of SATD in source-code corpora, where workaround and requirement debts have been found to be the most prevalent as well~\cite{bavota_large-scale_2016,maldonado_using_2017,potdar_exploratory_2014,liu_is_2020}.
One major departure from existing SATD literature is the high occurrence of \textit{Defect/Workaround} in our dataset. 
\textit{Defect/Workaround} is a type of Dockerfile-specific SATD that highlights a workaround due to bugs in external systems~\cite{azuma_empirical_2022}.
This reflects the nature of Dockerfiles that have to compensate for the behavior of external base images and upstream dependencies that are outside the developer's control.

\vspace{0.5em}\noindent
\textbf{A substantial share of admitted SATD comments in Dockerfiles are coupled with non-Dockerfile artifacts, and repaid SATD comments have even higher coupling rates.}
Figure~\ref{fig:coupling-rates}(a) shows that about 27\% of SATD admission events are coupled, whereas roughly 73\% are isolated.
In contrast, Figure~\ref{fig:coupling-rates}(b) shows 40\% of repayment events are coupled and about 60\% are isolated.
Methodologically, this result validates the co-evolutionary perspective of our study: a Dockerfile-only view may be misleading when studying SATD removal, because repayment is frequently conditioned on changes elsewhere in the repository.
This finding challenges a tacit assumption in the SATD repayment literature, which has predominantly measured repayment by examining edits \emph{within} the same file or module where the debt was originally admitted~\cite{mastropaolo_towards_2024,gao_automating_2021,gao_automating_2024}.
The substantially higher coupling rate at repayment time suggests that the resolution of IaC debt is intrinsically cross-artifact, consistent with broader co-evolution evidence showing that Dockerfile changes and application changes are strongly correlated~\cite{wu_dockerfile_2020,jiang_co-evolution_2015}.
Furthermore, prior IaC quality studies have treated Dockerfile smells and debt as self-contained issues amenable to local refactoring~\cite{durieux_empirical_2024,rosa_fixing_2024}; our results imply that automated repair tools targeting Dockerfiles in isolation systematically miss the external preconditions required for a large fraction of repayments.

\begin{figure}[t] 
    \centering
    \includegraphics[width=\textwidth, trim={.5cm .5cm .5cm .25cm}]{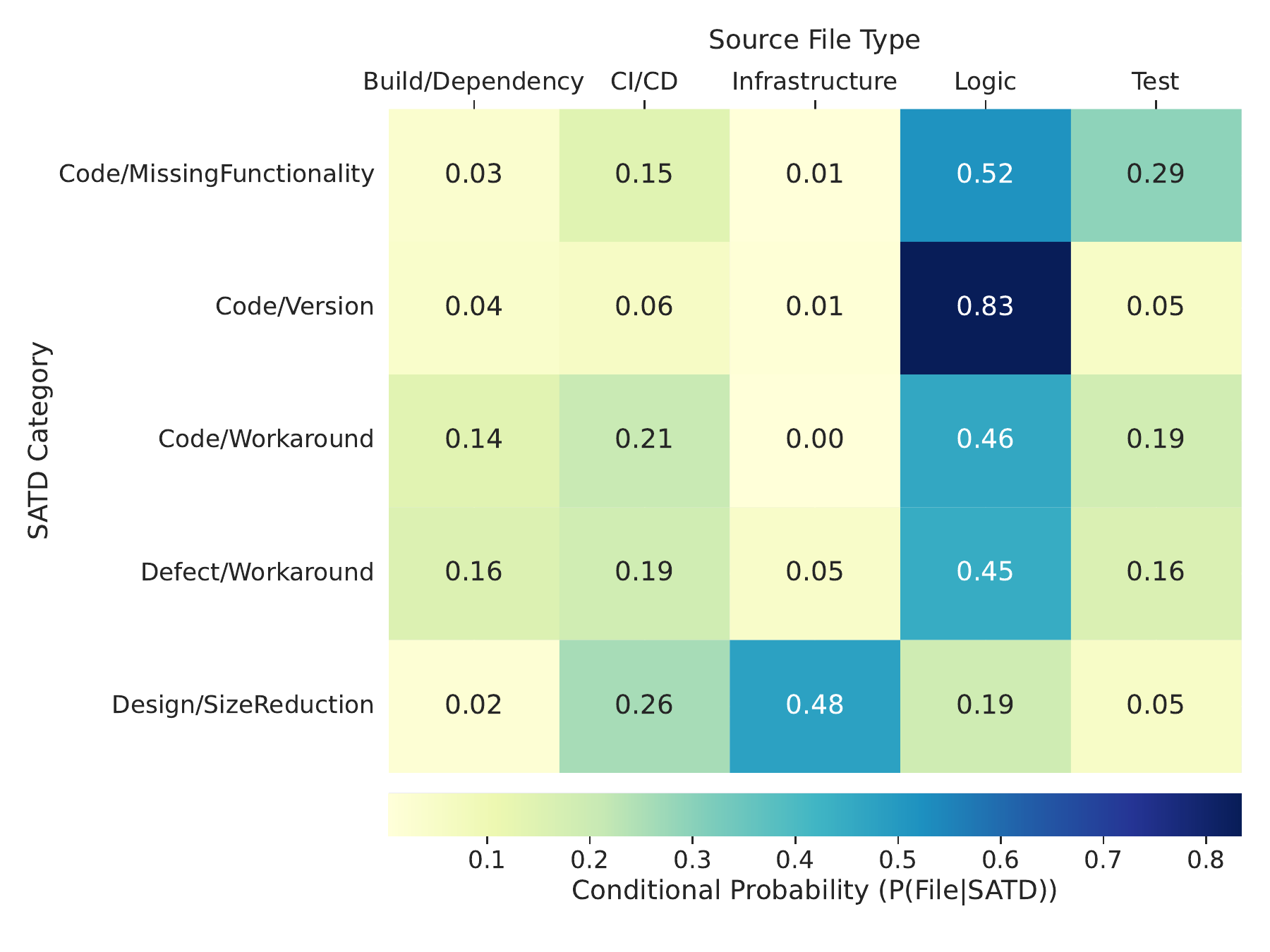} 
    \caption{Heatmap of conditional probability scores for the top-5 most prevalent SATD subtypes vs. co-changed file types.}
    \vspace{-1.5em}
    \label{fig:associations}
\end{figure}

\begin{table*}[tpb]
\centering
\caption{Correlation analysis between SATD subtypes and co-changed source file types sorted by Phi which denotes the binary correlation coefficient. Q-values are Benjamini-Hochberg adjusted Fisher exact test p-values.}
\label{tab:satd_filetype_correlation}
\begin{tabular}{|l|l|r|r|r|}
\hline
\textbf{SATD Subtype} & \textbf{File Type} & \textbf{$N$} & \textbf{$\phi$} & \textbf{$q$} \\ \hline
Process/Review & CI/CD & 10 & 0.157 & 0.009 \\ \hline
Process/Review & Test & 6 & 0.123 & 0.073 \\ \hline
Design/SizeReduction & CI/CD & 13 & 0.115 & 0.064 \\ \hline
Code/Version & CI/CD & 10 & 0.089 & 0.233 \\ \hline
Code/Workaround & CI/CD & 26 & -0.084 & 0.203 \\ \hline
Process/Deployment & Test & 3 & 0.074 & 0.547 \\ \hline
Defect/Workaround & CI/CD & 23 & -0.071 & 0.345 \\ \hline
Defect/Workaround & Infrastructure & 6 & 0.064 & 0.664 \\ \hline
Code/MissingFunctionality & CI/CD & 25 & 0.058 & 0.665 \\ \hline
Code/Version & Logic & 5 & 0.042 & 1.000 \\ \hline
Code/Workaround & Test & 17 & -0.042 & 1.000 \\ \hline
Code/Workaround & Build/Dependency & 23 & 0.042 & 1.000 \\ \hline
\end{tabular}
\end{table*}

\vspace{0.5em}\noindent
\textbf{Code-related SATD subtypes are predominantly co-occurring with application logic files.}
As shown in Figure~\ref{fig:associations}, \textit{Code/Version} has the highest conditional probability
with \textit{Logic} files ($P(\mathit{Logic}\mid\mathit{Code/Version})=0.83$), followed by
\textit{Code/MissingFunctionality}, \textit{Code/Workaround}, and \textit{Defect/Workaround}
($0.52$, $0.46$, and $0.45$, respectively).
Due to these relatively strong conditional probabilities, we can expect logic files to co-change in the
same commit where code-related SATD is admitted in a Dockerfile,
However, the corresponding $\phi$ coefficients for code-related SATD with co-changed \textit{Logic} files
are weak (e.g., $\phi=0.042$ for \textit{Code/Version}-\textit{Logic}).
Also, none of the SATD subtype-file type pairs reach statistical significance after Benjamini-Hochberg correction.
This suggests the binary correlation between SATD subtypes and file types is inconclusive even though co-change is frequent.
Therefore, Dockerfile comments about hacks, temporary workarounds, or missing features may
reflect broader architectural limitations rather than narrow coupling with a particular file type.
This finding does not contradict the IaC co-evolution literature, which has established
that source code is the primary driver of IaC file change~\cite{wu_dockerfile_2020,jiang_co-evolution_2015}.

\vspace{0.5em}\noindent
\textbf{Design-related SATD is positively correlated with CI/CD artifacts.}
Table~\ref{tab:satd_filetype_correlation} shows that \textit{Design/SizeReduction} exhibits
a statistically marginal positive correlation with \textit{CI/CD} files
($\phi=0.115$, $q=0.064$), and also has the highest conditional probability with
\textit{Infrastructure} files among all subtypes
($P(\mathit{Infrastructure}\mid\mathit{Design/SizeReduction})=0.48$).
Together, these results suggest that image-size and other image-optimization debt in Dockerfiles
tends to co-evolve with deployment architecture and integration pipeline configurations,
rather than arising from purely localized Dockerfile changes.
The CI/CD correlation, while not reaching the conventional $q<0.05$ threshold, is the
second-largest $\phi$ value in our dataset and reflects the practical impact of
build constraints on integration and deployment processes.
The only statistically significant correlation in our entire dataset is
\textit{Process/Review}-\textit{CI/CD} ($\phi=0.157$, $q=0.009$),
which indicates that process-oriented debt---such as comments flagging inadequate
review or approval steps---is reliably co-committed with CI/CD pipeline files.
Prior Dockerfile quality research has mainly studied image size as an aspect of
Dockerfile that can be optimized through localized instruction
changes~\cite{durieux_parfum_2023,bui_dockercleaner_2023}.
Our correlation results suggest a different mechanism: size-reduction decisions
in practice co-occur with changes to surrounding infrastructure and CI/CD
configurations, pointing to a systemic rather than purely local remediation pattern.

\begin{tcolorbox}[title=Summary of RQ1,colback=gray!5,colframe=black!65]
Dockerfile SATD is dominated by a code-related subtypes such as \textit{Code/Workaround}, \textit{Defect/Workaround}, and \textit{Code/MissingFunctionality}. 
Across all subtypes, non-trivial proportions of SATD events are coupled with source code changes in both admission and repayment scenarios.
Correlation analysis shows that coupling sources are subtype-specific: code-related SATD subtypes co-occur most frequently with co-change in \textit{Logic} files, 
whereas size optimization-related SATD shows a statistically marginal positive correlation with \textit{CI/CD} files.
\textit{Process/Review} is the only subtype with a statistically significant correlation, 
found with \textit{CI/CD} files.
\end{tcolorbox}

\section{Survival Analysis (RQ2)}
\label{sec:results-rq2}

\subsection{Motivation}

Although RQ1 quantifies how often Dockerfile SATD admissions and repayments are coupled with non-Dockerfile artifacts, frequency alone does not reveal whether such coupling is associated with different dynamics in which developers repay the debt.
For instance, coupled SATD instances may exhibit distinct repayment dynamics from isolated SATD instances.
More specifically, both the subtype of an SATD instance and its coupling status (\emph{coupled} or \emph{isolated}) can influence the dynamics of its repayment.
Analyzing SATD repayments along these two dimensions provides evidence on whether some forms of Dockerfile SATD should be expected to persist longer, require coordination across teams, or benefit from different debt management strategies.
Thus, for RQ2, we investigate whether \emph{coupled} and \emph{isolated} Dockerfile SATD instances differ in their \emph{time-to-repayment}.

\subsection{Approach}

\begin{figure}[t] 
    \centering
    \includegraphics[width=\textwidth, trim={.15cm .25cm .75cm .25cm}, clip]{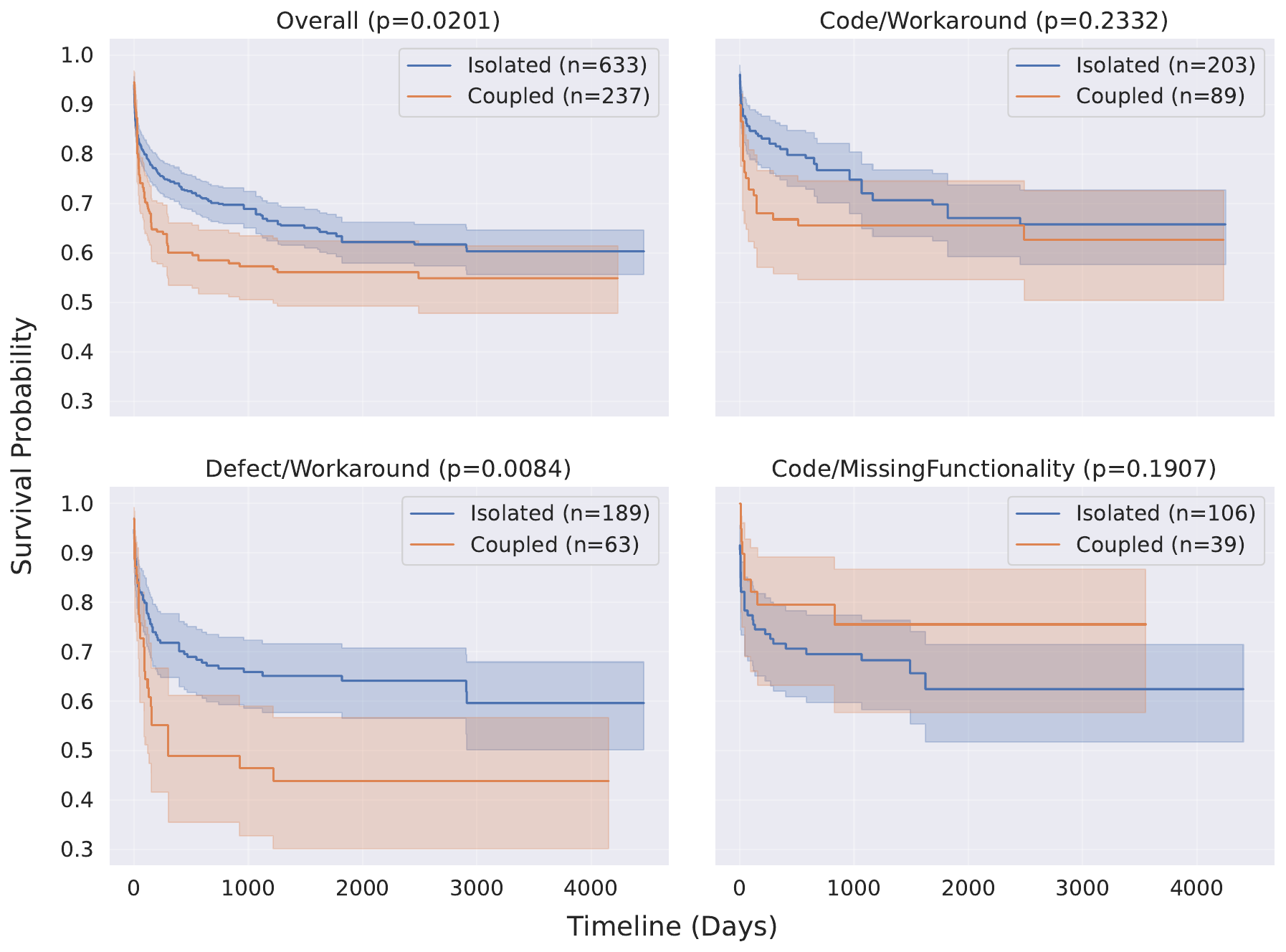} 
    \caption{Kaplan--Meier survival curves comparing isolated and coupled SATD instances overall and for the three most frequent SATD subtypes. Lower curves indicate a lower probability of remaining unrepaid over time. Shaded areas denote confidence intervals.}
    \label{fig:rq2_survival_grid}
    \vspace{-0.5em}
\end{figure}

Followin previous empirical studies in SATD repayment~\cite{liu_exploratory_2021,muse_fixme_2022,zampetti_was_2018,bernardo_impact_2023}, we model SATD repayment as a survival problem~\cite{kalbfleisch_statistical_2002}, where each SATD admission instance enters the risk set at the time it is introduced and exits when it is repaid. Instances not repaid by the end of the observation window are treated as right-censored. We compare two groups: \emph{Isolated} SATD events, whose admission/repayment commits have no manually identified semantically linked co-changed non-Dockerfile files, and \emph{Coupled} SATD events, for which at least one co-changed file was manually judged to be semantically linked.
\begin{enumerate}
    \item First, we examine whether Kaplan--Meier survival curves~\cite{kaplan_nonparametric_1958} visually diverge, indicating different repayment dynamics over time.
Figure~\ref{fig:rq2_survival_grid} shows the survival curves for all SATD instances combined and for the three most common SATD subtypes. 
    \item Then, we use the log-rank test~\cite{peto_asymptotically_1972} to assess whether these differences are statistically distinguishable.
Table~\ref{tab:survival_logrank} reports the corresponding log-rank tests, sample sizes, and approximate statistical power. 
    \item Finally, we inspect median time-to-repayment (TTR) values~\cite{kalbfleisch_statistical_2002} among repaid instances to characterize the practical direction of the difference for those instances that are eventually repaid.
Table~\ref{tab:ttr_comparison} reports median TTR in days among repaid instances only, separately for isolated and coupled groups.
\end{enumerate}

\subsection{Results}

\noindent
\textbf{Overall, coupled Dockerfile SATD instances are repaid faster than isolated SATD.} 
Top-left of Figure~\ref{fig:rq2_survival_grid} shows the survival curve of coupled SATD remains consistently below the curve of isolated SATD for all SATD subtypes combined.
This difference is statistically significant for the two population ($\chi^2=5.4067$, $p=0.0201$) as shown in Table~\ref{tab:survival_logrank}. 
Moreover, Table~\ref{tab:ttr_comparison} shows a slightly shorter median TTR for coupled instances than isolated ones ($39$ vs.\ $41$ days). 
This result contradicts our initial intuition and prevailing assumption that coupled SATD events might be harder and 
take longer to repay because they represent broader and more complex architectural limitations and compromises.
Instead, our finding indicates that developers may dedicate more effort and priority toward resolving more complex technical debt instances.

To investigate why coupled Dockerfile SATD in general is repaid faster, we conducted a hypothesis test
about the structural difference between coupled SATD events and the isolated ones.
More specifically, we use the Mann-Whitney U test~\cite{conroy_what_2012} to confirm statistical difference ($p=0.0127$) in the number of co-changed files
between the coupled and isolated SATD events.
This is consistent with findings from adjacent fields of bug-triaging~\cite{valdivia_garcia_characterizing_2014} and issue-tracking~\cite{ren_empirical_2020} 
where issues with broader system impact tend to be assigned higher priority and attract more attention due to their blocking nature that delay other works.

\vspace{0.5em}\noindent
\textbf{The SATD subtype with most difference in survival between coupled and isolated events is \emph{Defect/Workaround}, where coupled SATD is significantly less persistent than isolated SATD.} 
The bottom-left of Figure~\ref{fig:rq2_survival_grid} shows the coupled curve drops steeply early in the timeline and stays well below the isolated curve throughout the observation window. 
Log-rank test shown in Table~\ref{tab:survival_logrank} confirms the statistical significance of this difference ($\chi^2=6.9559$, $p=0.0084$), with the highest power of all subtype analyses ($0.7509$). 
On the other hand, Table~\ref{tab:ttr_comparison} shows that isolated \emph{Defect/Workaround} SATD has median TTR of $47$ days, compared with $51$ days for coupled instances.
This minor reversal in median indicates that the key distinction is not necessarily that every coupled workaround is repaid faster, but that coupled workarounds are less likely to remain indefinitely unrepaid.
This is in line with prior works that showed that workarounds for bugs in upstream libraries tend to accumulate as long-lived debt~\cite{maipradit_wait_2020,xiao_characterizing_2022}.
However, the bug is being actively tracked by maintainers, which in turn increases the probability that the workaround will eventually be retired rather than left indefinitely.
This aligns with the finding from the aforementioned hypothesis test on structural difference that confirmed coupled SATDs are more often repaid, 
as they represent coordinated remediation efforts that comprises higher-priority bug fixes observed in open-source maintenance
studies~\cite{valdivia_garcia_characterizing_2014,ren_empirical_2020}.

\begin{table}[t]
\centering
\caption{Survival Analysis and Log-Rank Statistical Power per SATD subtype, sorted by statistical significance (p-value) in descending order.}
\label{tab:survival_logrank}
\begin{tabular}{|l|c|c|c|c|c|}
\hline
\textbf{SATD Category} & \textbf{\# Iso.} & \textbf{\# Cpl.} & $\boldsymbol{\chi^2}$ & \textbf{p-value} & \textbf{Power} \\ \hline
  Defect/Workaround & 189 & 63 & 6.9559 & 0.0084 & 0.7509 \\ \hline
  Code/Version & 17 & 20 & 5.0185 & 0.0251 & 0.6104 \\ \hline
  Code/BaseImage & 18 & 4 & 1.9429 & 0.1634 & 0.2857 \\ \hline
  Code/MissingFunctionality & 106 & 39 & 1.7122 & 0.1907 & 0.2574 \\ \hline
  Test/IntegrityCheck & 13 & 5 & 1.5687 & 0.2104 & 0.2396 \\ \hline
  Design/SizeReduction & 41 & 3 & 1.4574 & 0.2273 & 0.2258 \\ \hline
  Code/Workaround & 203 & 89 & 1.4214 & 0.2332 & 0.2213 \\ \hline
  Test/ImprovementForTest & 4 & 2 & 1.25 & 0.2636 & 0.1999 \\ \hline
  Unclassifiable & 13 & 4 & 0.4865 & 0.4855 & 0.1034 \\ \hline
  Process/Review & 17 & 5 & 0.4447 & 0.5049 & 0.098 \\ \hline
  Process/Deployment & 10 & 3 & 0.3 & 0.5839 & 0.0789 \\ \hline
  Defect/LatentBug & 2 & 0 & - & - & - \\ \hline\hline
  Overall & 633 & 237 & 5.4067 & 0.0201 & 0.6425 \\ \hline
\end{tabular}
\end{table}

\vspace{0.5em}
\noindent
\textbf{\emph{Code/Workaround} shows a similar directional pattern as \emph{Defect/Workaround}, but the evidence is not strong enough to claim a reliable difference.} In Figure~\ref{fig:rq2_survival_grid} (top-right), the coupled curve is generally below the isolated curve early on, suggesting that coupled instances may be repaid somewhat faster.
Moreover, Table~\ref{tab:ttr_comparison} shows that among repaid instances, coupled \emph{Code/Workaround} SATD has a much shorter median TTR than isolated SATD (27 vs.\ 84 days).
This large difference in medians suggests that the difference remains consistent within the subtype.
However, Table~\ref{tab:survival_logrank} shows that the survival difference is not statistically significant ($p=0.2332$), and the estimated power is low (0.2213).
Thus, the data suggest a potentially meaningful practical effect, but one that our existing sample cannot establish with confidence.
Despite this lack of statistical significance, the large gap in median TTR ($27$ vs.\ $84$ days) provides an alternative perspective to the survival analysis, following
prior empirical studies SATD repayment~\cite{maldonado_empirical_2017,bhatia_empirical_nodate,muse_fixme_2022}.

\vspace{0.5em}\noindent
\textbf{\emph{Code/MissingFunctionality} exhibits the opposite trend from worka-round-related SATD subtypes: coupled SATD appears more persistent than isolated SATD.}
Bottom-right of Figure~\ref{fig:rq2_survival_grid} shows the coupled survival curve above the isolated one throughout the timeline, indicating a higher probability of remaining unrepaid.
Median TTR statistics in Table~\ref{tab:ttr_comparison} reinforces this observation, as repaid coupled instances have a longer median TTR than isolated instances ($42$ vs.\ $10$ days). 
This could indicate that SATD in Dockerfile caused by \emph{missing functionality} may require non-trivial implementation in the source code that require more substantial effort compared to simple cleanup and refactoring.
However, as with \emph{Code/Workaround}, the log-rank test shown in Table~\ref{tab:survival_logrank} is not statistically significant ($p=0.1907$). 
This pattern is consistent with findings in the SATD literature regarding requirement debt~\cite{maldonado_using_2017,potdar_exploratory_2014,bavota_large-scale_2016}, where missin feature debt has been identified as the most persistent and costliest category because its repayment is contingent on implementation work rather than simple cleanup.

\begin{table}[t]
\centering
\caption{Time-to-Repayment (TTR) in Days Comparison for Repaid Instances, sorted by Median TTR for Coupled SATD instances in ascending order.}
\label{tab:ttr_comparison}
\begin{tabular}{|l|c|c|c|c|}
\hline
\textbf{\begin{tabular}{@{}c@{}}SATD Category\end{tabular}} & \textbf{\begin{tabular}{@{}c@{}}\# Iso.\\Repaid\end{tabular}} & \textbf{\begin{tabular}{@{}c@{}}Iso. Median\\TTR\end{tabular}} & \textbf{\begin{tabular}{@{}c@{}}\# Cpl.\\Repaid\end{tabular}} & \textbf{\begin{tabular}{@{}c@{}}Cpl. Median\\TTR\end{tabular}} \\ \hline
  Unclassifiable & 4 & 1.0 & 2 & 18.0 \\ \hline
  Code/Workaround & 59 & 84.0 & 31 & 27.0 \\ \hline
  Code/Version & 10 & 121.5 & 18 & 27.0 \\ \hline
  Code/MissingFunctionality & 37 & 10.0 & 9 & 42.0 \\ \hline
  Defect/Workaround & 67 & 47.0 & 32 & 51.0 \\ \hline
  Design/SizeReduction & 13 & 21.0 & 2 & 58.5 \\ \hline
  Code/BaseImage & 12 & 30.5 & 1 & 253.0 \\ \hline
  Process/Review & 11 & 8.0 & 3 & 563.0 \\ \hline
  Test/IntegrityCheck & 3 & 415.0 & 2 & 636.5 \\ \hline
  Process/Deployment & 1 & 27.0 & 0 & - \\ \hline
  Test/ImprovementForTest & 4 & 1137.5 & 0 & - \\ \hline
  Defect/LatentBug & 0 & - & 0 & - \\ \hline\hline
  Overall & 221 & 41.0 & 100 & 39.0 \\ \hline
\end{tabular}
\end{table}

\noindent
\textbf{Other subtypes and statistical strength.}
Beyond the three most prevalent SATD subtypes, \emph{Code/Version} is the only subtype to exhibit statistical significance in survival difference ($p=0.0251$) where coupled instances more likely to be repaid than isolated ones.
However, the caveat is that this finding rests on a small sample (\#iso=$17$, \#cpl=$20$) with an estimated power of only $0.6104$.
Several other subtypes (\emph{Process/Review} and \emph{Code/BaseImage}) exhibit large descriptive gaps in median TTR, but are too sparse to support strong inferential claims.
This general limitation in statistical power for subtype-level analyses is natural result of the long tail of Dockerfile SATD categories first presented by Azuma et al.~\cite{azuma_empirical_2022} and later corroborated by us in Section~\ref{subsec:satd-classification}.
This implies that a purely quantitative lens does not suffice for explaining the nature of Dockerfile SATD and its co-evolution and thus, warrants a qualitative study follow-up for all the SATD subtypes in RQ3 (Section~\ref{sec:results-rq3}).

\begin{tcolorbox}[title=Summary of RQ2,colback=gray!5,colframe=black!65]
Coupled Dockerfile SATD instances are repaid faster than isolated ones overall ($p=0.02$), contradicting our initial intuition that cross-artifact debt incurs greater costs in terms of repayment time.
In particular, with \emph{Defect/Workaround}, coupled instances are significantly less likely to remain unrepaid ($p=0.0084$).
Our further quantitative analysis reveals that coupled SATD events are repaid with statistically higher number of co-changed files ($p=0.0127$), in line with
prior findings where co-evolution means active bug tracking and issue prioritization.
While \emph{Code/Workaround} exhibits a large gap in median TTR ($27$ vs.\ $84$ days), the effect is not statistically significant.
Conversely, \emph{Code/MissingFunctionality} coupled SATD appears more persistent than isolated SATD, consistent with requirement debt that is contingent on substantial implementation work rather than simple cleanup.
\end{tcolorbox}

\section{SATD-source code Co-evolution Patterns (RQ3)}
\label{sec:results-rq3}

\subsection{Motivation}

As presented in previous sections, Dockerfile SATD comments have non-trivial coupling rates both during admission and repayment.
The existing work that taxonomizes SATD in Dockerfile~\cite{azuma_empirical_2022} only looks at the content inside the comment (e.g., \textit{Defect},
\textit{Workaround}) and surrounding code chunk, but they do not study the \emph{cross-artifact} reasons
that caused the debt to be admitted or that eventually enabled its repayment.
Without this perspective, both practitioners and researchers cannot answer actionable questions such as: \textit{``What kind of change in our application code is most likely to introduce a new Dockerfile SATD?''} or \textit{``What kind of code evolution is a prerequisite for retiring a suboptimal Dockerfile workaround?''}

To fill this gap, we conduct a qualitative study that taxonomizes
(a)~the \textbf{admission triggers} which are cross-artifact changes that
necessitate the introduction of a Dockerfile SATD comment, and 
(b)~the \textbf{repayment requisites} which are cross-artifact changes that
enable the removal of a Dockerfile SATD comment.
These two taxonomies collectively answer RQ3 and complement the quantitative coupling measurements reported for RQ1 and RQ2.

\subsection{Approach}
\label{sec:opencode-approach}

Following prior taxonomy studies in empirical software engineering~\cite{cai_fortifying_2026,openja_empirical_2024,watanabe_use_2026}, 
we conduct open-coding and axial-coding to characterize the source code-side co-evolution
in the SATD events in our dataset.

\paragraph{Co-evolution dataset.}
Our SATD coupling annotation in Section~\ref{subsec:event-coupling} produced a set of $411$
\emph{coupled} SATD events where there is at least
one co-changed non-Dockerfile file was manually labelled as \textsc{semantically-linked}.
Each SATD event is represented by the schema described in Table~\ref{tab:annotation-schema} of
Section~\ref{subsec:data-annotation}, containing the SATD comment,
its Dockerfile context, the commit message, and the diff snippets of
all semantically linked co-changed files.

\paragraph{Open coding.}
In the pilot round, we randomly sample a subset of 15 events from each SATD subtype.
If a subtype has less than 15 events, we include all of the events for the subtype in this iteration.
Consequently, we end up with a pilot set of 82 admission events and 82 repayment events
(164 units in total).
The first three authors independently analyzed the pilot set where for each SATD event, 
they read the full context of the SATD event, (the SATD comment
text, the surrounding Dockerfile instructions, the commit message, and the
diff snippets of semantically-linked co-changed files), and drafted a free-text
\texttt{category} label and corresponding \texttt{rationale} that captures
(i)~what changed in the source code file, and (ii)~why that change
caused the Dockerfile SATD to be admitted or enabled it to be repaid, 
After the pilot round, a synchronization meeting was held: the annotators discussed
the initial codes and disagreements, and produced a preliminary
fine-grained codebook that comprises source codce change patterns observed in the 
SATD events both admission and repayment events.
Using the preliminary codebook, the annotators independently applied it
to the remaining $249$ coupled events in the co-evolution dataset.
An adjudication meeting was held to resolve labeling conflicts and
to refine the codebook.

\paragraph{Axial coding.}
After the fine-grained codebook stabilized, the first three authors
discussed the categories and grouped the finer-grained codes into higher-level themes.
Each higher-level theme was assigned a concise name, a definition, 
with representative examples drawn from the dataset for each fine-grained codes, 
and highlight of the common theme across those examples.

\paragraph{Validation.}
The draft taxonomies were presented to the remaining co-authors in a
face-to-face review session. 
For each category, all the co-authors discussed the category name,
definition, and representative examples.
The first three authors answered the questions raised by the remaining co-authors, 
and rearranged themes for some of the fine-grained codes, and clarified cases where
the boundaries were ambiguous and causing confusions.
The taxonomies were revised iteratively until consensus was reached among all
authors.

\subsection{Results}

\begin{figure}[t]
  \centering
  \includegraphics[width=\linewidth]{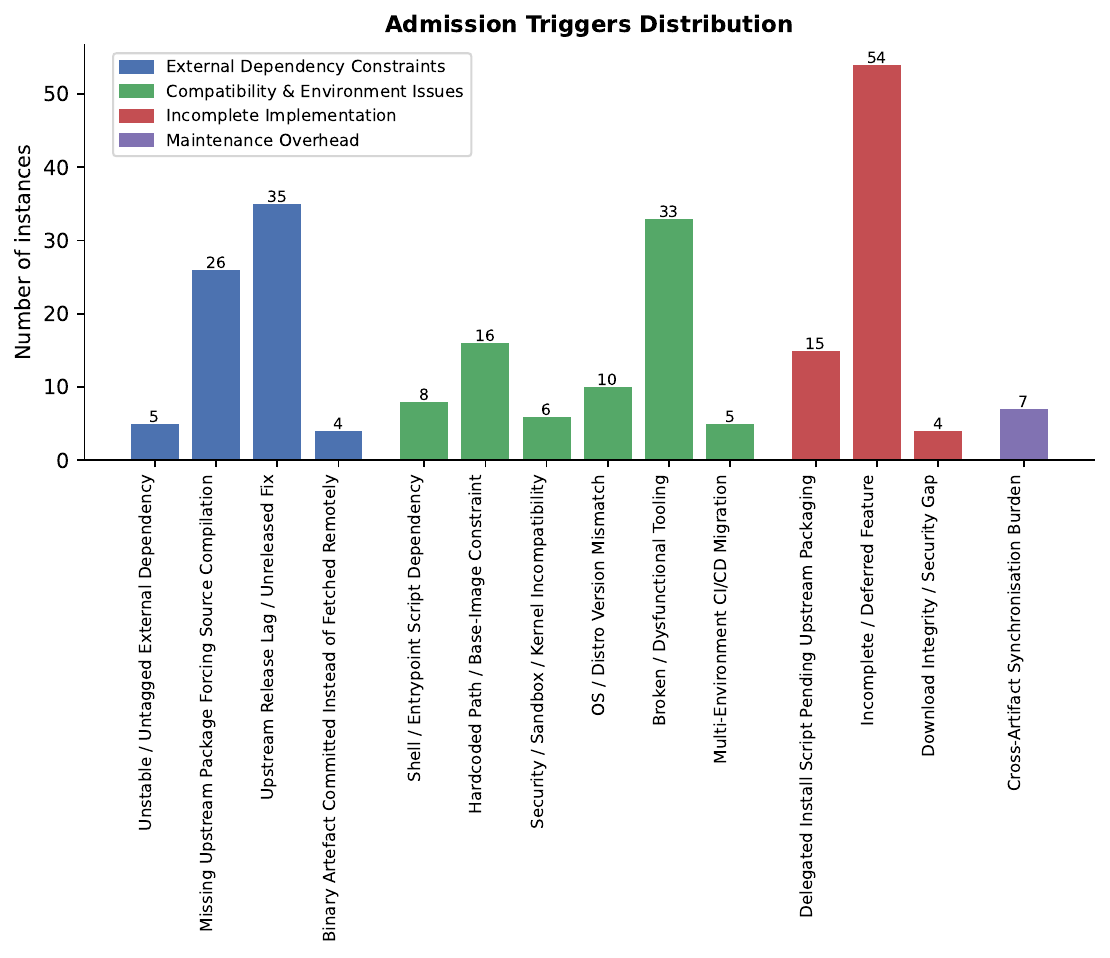}
  \caption{Taxonomy of \emph{admission triggers} for coupled
           Dockerfile SATD, clustered into four overarching themes.}
  \label{fig:rq3_admission_grouped}
\end{figure}

\begin{figure}[t]
  \centering
  \includegraphics[width=\linewidth]{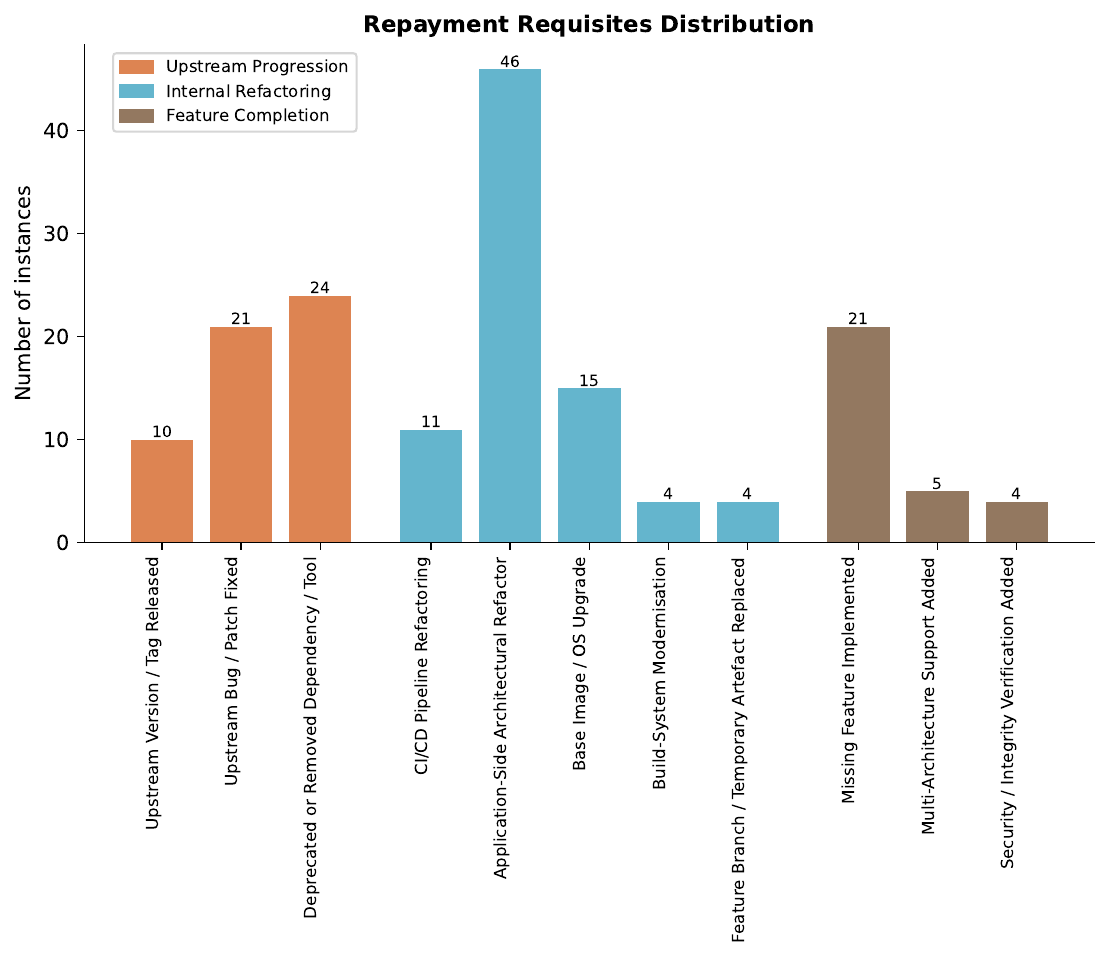}
  \caption{Taxonomy of \emph{repayment requisites} for coupled
           Dockerfile SATD, clustered into 3 overarching themes.}
  \label{fig:rq3_repayment_grouped}
\end{figure}

Figures~\ref{fig:rq3_admission_grouped} and~\ref{fig:rq3_repayment_grouped}
show the taxonomies of SATD admission triggers, and repayment requisites 
discovered via the open- and axial-coding process.
We identified $14$ types of admission triggers clustered under $4$ overarching themes, 
and $11$ types of repayment requisites under $3$ overarching themes.
Example SATD instances discussed below can be all found in full context 
(i.e., SATD comment, commit message, surrounding code chunks, diff snippets of co-changed files, etc...) 
in our publicly available dataset~\cite{minn_dockerfile_nodate}.

\subsubsection{Admission Triggers}

\paragraph{Admission Theme 1 – External Dependency Constraints.}
This cluster comprises $80$ instances across $4$ categories of SATD admission triggers
that arise outside the developer's direct control.

\vspace{0.5em}\noindent
\textbf{Unstable / Untagged External Dependency.}
Package manifests and dependency configurations could include 
SemVer tags~\cite{preston-werner_semantic_nodate} to not yet stable versions. 
Therefore, developers have to delete lock-files, and use temporary/placeholder 
identifiers as workarounds in the Dockerfiles.
For example, a \texttt{package.json} included direct GitHub references for
\texttt{lnd-engine} and \texttt{grpc-methods} without proper version pins,
and the Dockerfile was forced to delete \texttt{package-lock.json} so that
its \texttt{npm install} command could re-resolve these packages dynamically~\footnote{row 4}.
admitted the technical debt: \emph{``hack for Ruby 2.7.3''} in the Dockerfile~\footnote{rows 41, 44}.


\vspace{0.5em}\noindent
\textbf{Missing Upstream Package Forcing Source Compilation.}
When a new dependency is not yet available inside the base image, 
developers are forced to do multiple source-compilation
as a workaround in the Dockerfile.
For example, \texttt{prometheus-client-c} v0.1.3 was unavailable in Alpine's package
index. This forces the developers to add in a dedicated \texttt{dist-libprom} build stage 
to compile it from source~\footnote{rows 10, 73, 74}.
Alongside the workaround, the developer stated in the SATD comment that it should be 
revisited \emph{``when Alpine packages the dependency''}.

    
\vspace{0.5em}\noindent
\textbf{Upstream Release Lag / Unreleased Fix.}
The upstream fix or file is available online, but not officially released yet.
Therefore, the developer have to pin 
hard-code pre-release branch names or raw commit hashes in the Dockerfile.
For example, the offical release tag for the upstream \texttt{valkey-search} repository  
does not contain the required fix when packaged together with Alpine Linux.
In this case, it is caused by the developer's decision to add Alpine Linux support, but the Alpine Linux image's pre-packaged \texttt{valkey-search} dependency does not contain the fix yet.
Therefore, the developer had to clone the upstream at the specific commit that contains the nessary fix~\footnote{rows 53-55, 60-63, 66},


\vspace{0.5em}\noindent
\textbf{Binary Artifact Committed Instead of Fetched Remotely.}
When the developer copies a binary directly into the repository as a workaround
rather than fetching it via \texttt{curl}, they are forced to admit 
a \texttt{TODO} comment inside the Dockerfile to switch to a proper 
remote download in the future~\footnote{row 2}.


\paragraph{Admission Theme 2 – Compatibility and Environment Issues.}
$78$ instances arise from mismatches between the container build or runtime
environment, and the assumptions in source code, or other configuration files.

\vspace{0.5em}\noindent
\textbf{Shell / Entrypoint Script Dependency.}
When startup scripts adopt Bash-specific features,
the developer must replace the default POSIX shell with Bash in the Dockerfile. 
Alongside the Bash macro, the developers also usually admit a ``hack'' to 
annotate this as necessary but a suboptimal configuration.
For example, after a co-changed \texttt{ros\_entrypoint.sh} adopted \texttt{\#!/bin/bash}
and relied on the \texttt{source} builtin, the Dockerfile was forced to
replace \texttt{/bin/sh}~\footnote{row 0}.
In another case, the developer added a new \texttt{xrdp/start.sh} startup script 
whose execution context depends on \texttt{PATH} variable, and thus was forced to 
do a redundant \texttt{PATH} export in the Dockerfile~\footnote{row 14}.

\vspace{0.5em}\noindent
\textbf{Hardcoded Path / Base Image Constraint.}
When the upstream base image assumes rigid path formats, the developer
is forced to use a workaround in Dockerfile to match the expected filesystem layout.
For example, the Hyperledger Fabric base image assumes the existence of \texttt{/opt/gopath} path.
Thus, the developer inserted a symlink (\texttt{ln -s \$GOPATH /opt/gopath}) as a workaround which 
is admitted as \emph{``hard-coded problem''}~\footnote{row 1}.
In another instance, OpenRV's custom PySide2 compilation was conflicting with system-installed Qt.
This required the developer to reinstall Qt into the \texttt{/tmp/qttemp} path as a workaround~\footnote{row 34}.

\vspace{0.5em}\noindent
\textbf{Security / Sandbox / Kernel Incompatibility.}
When the Docker build environment comes with security constraints that impede building or running of the application, 
the developer is forced to use workarounds to circumvent the security constraints.
For example, two independent commits describe FUSE installation as requiring manual
\texttt{.deb} repackaging because Docker's overlay filesystem does not support
\texttt{mknod} during build~\footnote{rows 37, 39}.
In another case, a \texttt{virtiofsd} Rust patch disables sandboxing and comments out
capability-enforcement errors, acknowledged as a \emph{``huge ugly hack that
barely works''}~\footnote{row 6}.

\vspace{0.5em}\noindent
\textbf{OS / Distro Version Mismatch.}
When there is OS or distribution version mismatches caused by base-image
upgrades in co-changed build scripts, the developer have to resolve the version conflicts using workarounds in the Dockerfiles.
For example, upgrading from Ubuntu 20.04 to 24.04 in a co-changed
\texttt{build\_openfoam.sh} introduced \texttt{apt-get} temporary-file failures.
This required the developer to add a manual directory cleanup workaround~\footnote{rows 33, 38}.

\vspace{0.5em}\noindent
\textbf{Broken / Dysfunctional Tooling.}
When a co-changed script or dependency introduces a bug in a tool
that the Dockerfile relies upon, the developer admit the workaround and
references the upstream issue.
For example, a co-changed \texttt{entrypoint.sh} used \texttt{tlmgr} to install 
TeX Live package.
However, the tool was not functional in the Alpine build context,
thus forcing the developer to admit a work-in-progress SATD comment~\footnote{row 5}.
Likewise, a bug in the Transifex generation process prevented pulling
\texttt{nutmeg.1} translation strings, and forced the developer to add a workaround with
a hardcoded \texttt{open-release/nutmeg.master} version in the Dockerfile~\footnote{row 36}.

\vspace{0.5em}\noindent
\textbf{Multi-Environment CI/CD Migration.}
When migrating the project to a new CI environment, 
the Dockerfiles in the project need to remain compatible with configuration value formats 
expected by both the new and old (legacy) CI environments.
For example, \texttt{ray-project/ray} project migrated from old CI environment, \texttt{civ1}, 
to the new CI environment, \texttt{civ2}, which expects the \texttt{requirements.txt} files to 
be under \texttt{python/} subdirectory unlike \texttt{civ1}.
This could cause exception when deploying in legacy \texttt{civ1} environment as the 
\texttt{python/} path is non-existent.
Thus, the developer had to "workaround" this exception in \texttt{civ1} by 
ignoring the missing sources~\footnote{row 7}. 

\paragraph{Admission Theme 3 – Incomplete Implementation.}
There are $73$ instances across $3$ categories of admission triggers 
indicating the deferral of feature- and security-related implementations, 
or packaging steps.

\vspace{0.5em}\noindent
\textbf{Delegated Install Script Pending Upstream Packaging.}
When upstream binary packaging infrastructure is not yet available, developers
introduce external shell scripts as placeholders and admit the debt explicitly.
For example, four variants of the CNTK image introduced a \texttt{install-cntk-docker.sh} 
script as a temporary proxy to handle proper binary distribution mechanism for 
Anaconda setup, which the developer could not implement immediately yet. 
Thus, the developer admitted \emph{``TODO: Implement via modifying Binary Drop
package''} in the Dockerfile~\footnote{rows 3, 11, 13, 40}.

\vspace{0.5em}\noindent
\textbf{Incomplete / Deferred Feature.}
Some features implemented in the source code and configuration files are 
finalized with corresponding commands inside the Dockerfile.
Hence, there are cases where those corresponding commands cannot be implemented 
inside the Dockerfiles by the developer yet due to a multitude of unforeseen reasons.
For example, in \texttt{tianon/dockerfiles} project, the developer added 
a stable variant of a new feature to the configuration, but was unable to 
conduct verification with non-\texttt{amd64} variant of dependencies.
Thus, the developer had to hardcode \texttt{amd64} parameter to the APT sources list
when installing dependencies in the Dockerfile~\footnote{rows 21}.
Thus, the developer notifies the maintainers about the technical debt

\vspace{0.5em}\noindent
\textbf{Download Integrity / Security Gap.}
As the developer introduces new externally fetch binaries to the project, 
the developer might also admit a TODO for a deferred integrity check command 
for the binaries inside the Dockerfile.
For example, when a co-changed \texttt{update.sh} script automated binary version tracking
from GitHub releases, the developer added a TODO noting that \texttt{.asc} and
\texttt{.sha256sum} verification were not yet implemented~\footnote{row 68}.
In another instance, a wiremock extension dependency was switched to a new Maven Central JAR without
checksum validation~\footnote{row 69}, and \texttt{js-yaml} downloads within a
MongoDB 8.0 upgrade lacked GPG verification despite the rest of the image using
rigorous PGP checks~\footnote{row 70}.

\vspace{1em}\noindent
\textit{Admission Theme 4 – Maintenance Overhead.}

\vspace{0.5em}\noindent
\textbf{Cross-Artifact Synchronisation Burden.}
During maintenance processes that involve refactoring or upgrading of an artifact,
the developer have to admit technical debt if they are not able to keep
the Dockerfiles consistent with the design decisions inside the artifact.
For example, installing \texttt{supervisord} for process management introduced 
complexities regarding zombie process issues requiring \texttt{tini}.
Thus, the developer had to admit in the Dockerfile a TODO comment referencing a
known GitHub issue about the interaction between the two tools~\footnote{row 30}.
In another case, during the migration \texttt{golangci-lint} from v1.46 to v1.49,
a \texttt{//nolint} directive in \texttt{cluster.go} could not be removed
until the linter was updated again~\footnote{row 42}.

\subsubsection{Repayment Requisites}

\paragraph{Repayment Theme 1 – Upstream Progression.}
$55$ repayment events are
enabled by events outside the project such as upstream version releases,
upstream bug fixes, or the outright removal of a deprecated dependency.

\vspace{0.5em}\noindent
\textbf{Upstream Version / Tag Released.}
When the SATD comment contains an explicit version condition, it can only be satisfied by
upstream version releases that are reflected by a version bump 
in a co-changed \texttt{versions.json}, or \texttt{manifest.yml}.
For example, an SATD comment regarding Perl 5.33.7 compatibility patch was repaid 
when a co-changed \texttt{build} file bumped \texttt{VERSION} 
from \texttt{0.9999} to \texttt{1.0.0}~\footnote{row 0}.
In another case, an Imagick PECL-incompatibility workaround in the Dockerfile 
was repaid after the release of \texttt{imagick-3.8.0} which was reflected in
co-changed \texttt{versions.json} and \texttt{update.sh} in the same commit~\footnote{rows 4, 14}.

\vspace{0.5em}\noindent
\textbf{Upstream Bug / Patch Fixed.}
When the bug identified inside the SATD comment has been fixed in the upstream dependency,
it is reflected in the changes in test files that where the developer can remove source 
code-side workarounds such as skip guards, and deferred assertions.
For example, in \texttt{dotcloud/docker} project, a workaround in Dockerfile where 
the developer used \texttt{iptables-legacy} instead of \texttt{nf\_tables} 
which was not supported by both the CI hosts and application code.
Only when the test file, \texttt{libnetwork/firewall\_linux\_test.go}, was updated with
a new runtime detection logic for \texttt{nf\_tables}, it indicates that the application code 
has been patched to support \texttt{nf\_tables}, thus allowing the workaround in the Dockerfile
to be removed as well~\footnote{row 7}.

\vspace{0.5em}\noindent
\textbf{Deprecated or Removed Dependency / Tool.}
The developers remove upstream tools (that may be deprecated)
thus, enabling the repayments of the technical debt that exists due to workaround required by 
the feature or tool that no longer exists in the codebase.
For example, the \texttt{ksonnet} workaround was repaid by the complete removal of ksonnet
support: installer scripts, documentation, and \texttt{hack/tool-versions.sh}
were all deleted in the same commit~\footnote{row 5}.
Also, three separate instances of an FPM-based Debian packaging workaround
(\emph{``TODO: replace FPM with minimal debhelper stuff''}) were repaid when
\texttt{hack/release.sh} removed the entire \texttt{release\_ubuntu()} function
and \texttt{RELEASE-CHECKLIST.md} was updated to use
\texttt{hack/make.sh build-deb}~\footnote{rows 33, 38, 43}.

\paragraph{Repayment Theme 2 – Internal Refactoring.}
There are $80$ repayment events across $5$ categories that we found to be
enabled by \emph{internal} factors such as: 
CI pipeline refactor, base image change, build system modernisation, or 
replacing a temporary artifact with a stable one.

\vspace{0.5em}\noindent
\textbf{CI/CD Pipeline Refactoring.}
The developer repays the SATD inside the Dockerfile by moving the
associated workaround out of the Dockerfile, and into a dedicated
build infrastructure. This is enabled by refactoring the pipeline into 
a multi-stage or multi-container build process.
For example, in \texttt{StefanScherer}\texttt{/dockerfiles-windows}, a PowerShell bug causing 
a specific MSI installation failure forced the developer into a workaround 
involving the installation of a full Windows SDK which is considered to be heavy.
To repay this hefty workaround, the developer used an intermediate Dockerfile solely 
for installing the SDK, and then extracted only the necessary library files 
into a temporary directory before building the main Dockerfile~\footnote{row 9}.

\vspace{0.5em}\noindent
\textbf{Application-Side Architectural Refactor.}
The developer performs the various application-side architectural refactoring such as 
modularisation, dependency restructuring, process-manager changes, and 
build-system consolidation, and consequently removes the corresponding workaround in the Dockerfile.
For example, the developer split the monolithic \texttt{install-binaries.sh} into per-binary
\texttt{*.installer} files that allowed the Dockerfile to remove a version
synchronisation-related SATD that explicitly referenced the old script~\footnote{rows 44, 57, 59}.
In another instance, the developer migrated from Hive-on-Hadoop to building Spark from source, and 
rendered an entire Guava-jar conflict workaround obsolete~\footnote{row 2}.

\vspace{0.5em}\noindent
\textbf{Base Image / OS Upgrade.}
The developer upgrades base image or OS to resolve version-specific workarounds 
caused by missing packages or runtime capabilities in previous distributions.
For example, both the \texttt{plpython3} Debian Sid hack and the Coturn
\texttt{mongo-c-driver} source-compilation workarounds were repaid by upgrading
the base image from Debian Buster to Bullseye, which made the required packages
available in official repositories~\footnote{rows 31, 37, 51, 52}.
Moreover, the co-changed CHANGELOG file explicitly notes the Debian version upgrade 
as the enabling factor in each of the cases.
In another case, upgrading from \texttt{tomcat:8.5-jre8} to an \texttt{s6-overlay}-compatible
base image enabled the introduction of proper service management
(\texttt{rootfs/etc/services.d/tomcat/run}), resolving a long-standing TODO
about rebasing~\footnote{row 48}.

\vspace{0.5em}\noindent
\textbf{Build-System Modernisation.}
Developers upgrade to a more modern build-system so that they can 
abstract the complex manual implementations away using a now capable framework.
For example, BuildKit's native \texttt{--platform} or \texttt{TARGETPLATFORM}
via \texttt{docker-bake.hcl} supports the cleaning of testing-only binaries 
from the main stage.
Thus, the manual hacks using \texttt{hack/make/cross} script and the associated \texttt{DOCKER\_CROSSPLATFORMS} variable inside the Dockerfile were now obsolete, 
and its associated SATD comments could now be removed~\footnote{rows 71, 72, 80, 81}.

\vspace{0.5em}\noindent
\textbf{Feature Branch / Temporary Artefact Replaced.}
The developer replaces a feature branch-specific temporary image with
the stable official Docker Hub image once the branch condition check inside 
the CI configuration file was removed.
For instance, in \texttt{baserow/baserow} project, the Dockerfile had been temporarily 
pinned to a GitLab CI registry image indicating a development branch.
The repayment of this SATD was enabled by the co-changed in \texttt{.gitlab-ci.yml} 
configuration file which removed the branch restriction check, and enabled the switch 
to the official \texttt{baserow/baserow:1.10.1} image~\footnote{row 53}.

\paragraph{Repayment Theme 3 – Feature Completion.}
There are $30$ instances across $3$ categories of SATD repayment events 
corresponding to the explicit implementation of
features or security practices such as implementation of missing features, 
adding multi-architectural support, and adding security/integrity verification.

\vspace{0.5em}\noindent
\textbf{Missing Feature Implemented.}
The developer adds previously absent source code, configuration, or test cases 
for the Dockerfile to invoke.
For example, in \texttt{tianon/dockerfiles} project, the 
\texttt{\# TODO copy/steal ``browser.sh'' stuff from slack} SATD comment was repaid at
the creation of \texttt{zoom/browser.sh} which implements the logic that uses Zenity library
to prompt the user for URL input~\footnote{row 26}.
In another case, in \texttt{mondoohq/installer} project, workaround in Dockerfile regarding 
over infeasible non-root container support 
was repaid when the CI pipeline introduced dedicated
\texttt{rootless} build targets that provided a clean separation between root and
rootless variants~\footnote{rows 39, 41}.

\vspace{0.5em}\noindent
\textbf{Multi-Architecture Support Added.}
The developer adds structured per-architecture metadata such as URLs, checksums, \texttt{dpkg} mappings
in version-tracking artifacts for the Dockerfile to reference.
This enables the repayment of technical debt regarding hard-coded configuration values in Dockerfile.
For example, in the \texttt{containerd} project, the developer extended \texttt{versions.json} 
with per-architecture entries for amd64, arm64, ppc64el, riscv64 and s390x, 
and added \texttt{deb\_arch} mapping function to \texttt{lib.jq}
This enable the repayment of \texttt{\# TODO multiple architectures} comment in the Dockerfile 
which now switches to dynamic \texttt{case} selection~\footnote{rows 19, 28, 77, 78}.
In another case, the developer added support for cross-compilation between \texttt{linux/386} and \texttt{linux/arm} 
in the \texttt{hack/release.sh} via \texttt{armel} handling and
added platform-agnostic build
tags in \texttt{pkg/netlink/netlink\_unsupported.go}~\footnote{rows 20, 24, 25, 27}.

\vspace{0.5em}\noindent
\textbf{Security / Integrity Verification Added.}
The developer added automated checksum or GPG infrastructure in co-changed
version-tracking scripts or release tooling.
This allows the repayment of download integrity-related SATD in the Dockerfile.
For example, the \texttt{\# TODO .sha256sum} for \texttt{containerd} was repaid when
\texttt{versions.sh} gained logic to fetch \texttt{.sha256sum} files
automatically, and \texttt{versions.json} was extended with per-architecture
SHA256 hashes~\footnote{rows 77, 78}.
In another instance, a GPG fingerprint TODO in an Apache Geode Dockerfile was resolved by converting
the static fingerprint to a \texttt{\$\$PLACEHOLDER\$\$} that
\texttt{promote\_rc.sh} fills via \texttt{sed} during release
promotion~\footnote{row 76}.

\vspace{.75em}
\begin{tcolorbox}[title=Summary of RQ3,colback=gray!5,colframe=black!65]
We present 14 types of admission triggers with 4 overarching themes 
(\textbf{external dependency constraints}, \textbf{compatibility and environment issues}, 
\textbf{incomplete implementation}, \textbf{maintenance overhead})
where external dependency constraints is the most common theme of admission triggers.
We also present 10 types of repayment requisites with 3 overarching themes
(\textbf{upstream progression}, \textbf{internal refactoring}, \textbf{feature completion})
where internal refactoring is the most common theme of repayment requisites.
\end{tcolorbox}

\section{Implications}
\label{sec:implications}

\subsection{Implications for Researchers}
\label{subsec:impl-researchers}

\vspace{0.5em}\noindent
\textbf{Extend automated SATD tools to consider cross-artifact relationships.}
Building on the preceding unit-of-analysis argument, existing automated SATD detection and repair approaches predominantly operate on individual files~\cite{mastropaolo_towards_2024,gao_automating_2024,gao_automating_2021}.
Our results indicate that such tools systematically miss the external preconditions for repayment.
In the SATD lifecycle example depicted by Figure~\ref{fig:motivation}, the version-related workaround SATD in the Dockerfile 
is caused by a Glide-related script somewhere else in the project.
From automated SATD annotation perspective, existing tool would not be able to detect the source code-side workaround that
implemented Glide module, and thus, would not be able to reflect this workaround inside the Dockerfile as well.
The repayment of this workaround also involves source code-side evolution that represents migration from Glide- to Go-based configuration files.
Automated this repayment process would require the tool to factor in such source code-side evolution to synthesize code changes
that encompasses the code chunks in the Dockerfile that is effected by migration that happened inside the source code.
Therefore, researchers building the next generation of automated tools for SATD detection, annotation, tracking, or repair should incorporate such cross-artifact signals 
such as code artifact co-change graphs, and diff snippets of co-changed alongside SATD comments for inputs.

\vspace{0.5em}\noindent
\textbf{Cross-artifact signals in other software engineering topics.}
Despite the prevalence of production-test co-evolution topic in software engineering~\cite{marsavina_studying_2014,vidacs_co-evolution_2018,levin_co-evolution_2017,wang_understanding_2021,sun_revisiting_2023,chi_reaccept_2025,li_cocoevo_2025,shimmi_leveraging_2022}, 
multiple topics in software engineering such as Dockerfile smells detection and repair~\cite{bui_dockercleaner_2023,durieux_parfum_2023,rosa_fixing_2024,shabani_dockerfile_2025}, Code-Comment Inconsistency (CCI) resolution~\cite{rong_code_2025,li_cocoevo_2025} are still limited in terms of their single-artifact scope.
Even though latest work in Dockerfile flakiness repair, FlakiDock~\cite{shabani_dockerfile_2025}, involves the inspection of exception messages during Docker image build process, 
they do not consider additional context from the source code-side artifacts that might be related to the exception, and posted modest performance improvement of only $0.58\%$ over the state-of-the-art, Parfum~\cite{durieux_parfum_2023}.
Hence, integrating cross-artifact signals to improve contextual roadblocks is a promising direction for researchers in other software engineering topics to investigate.

\subsection{Implications for Developers}
\label{subsec:impl-developers}

\vspace{0.5em}\noindent
\textbf{Annotate SATD with explicit cross-artifact context at admission time.}
The admission trigger taxonomy presented in RQ3 (Section~\ref{sec:results-rq3}) shows that most Dockerfile SATD originates from external artifacts/projects such as incomplete implementation, upstream release lag, and broken toolings.
Therefore, developers include explicit cross-artifact context in the SATD comment.
For example, including the link to the GitHub issue shows the status of technical debt caused by defects in the upstream dependency.
The paths to the related source or manifest files that caused the workaround allows maintainers to readily access the related files during the SATD repayment.
Such additional context also reduce the risk of persistent SATD that causes blockages to resolving issues and other technical debt~\cite{martini_danger_2015,valdivia_garcia_characterizing_2014,ren_empirical_2020}.

\vspace{0.5em}\noindent
\textbf{Extend upstream release monitoring tools to automate Dockerfile maintenance.}
In the admission trigger category of \textit{upstream release lag} described in RQ3 (Section~\ref{sec:results-rq3}), developers usually pin commit hashes or pre-release tags when the upstream dependency has not released the feature or bug fix.
Thus, this technical debt is simply waiting for the release of that feature or bug fix.
This means that linking the automated upstream dependency monitoring tools such as Dependabot~\cite{noauthor_dependabot_nodate} and Renovate~\cite{noauthor_dependency_nodate} can be extended to track dependency-related SATD in the project, and apply pre-written modifications to the Dockerfile to automatically repay SATD.

\section{Threats to Validity}
\label{subsec:threats}

\subsection{Internal validity.}

\paragraph{Keyword-based SATD candidate extraction.}
The first threat to internal validity is the use of keywords such as \texttt{TODO}, \texttt{FIXME}, \texttt{HACK}, and \texttt{WORKAROUND} to extract SATD candidate comments that can miss out actual SATD comments that do not contain those keywords.
However, this approach is following prior works in empirical studies about SATD both in Dockerfile domain~\cite{azuma_empirical_2022} and outside of it~\cite{guo_how_2021}.
Moreover, the completeness of this keyword-based approach has already been statistically validated by Azuma et al.~\cite{azuma_empirical_2022} that only $1.83\%$ of comments that do not contain SATD keywords are SATD.

\paragraph{Inter-rater disagreement during manual annotation tasks.}
We mitigate this risk with the three-phase protocol described in Section~\ref{subsec:data-annotation}: a pilot phase for codebook refinement, a reliability phase on a shared subset of $100$ instances per task, and a production phase with adjudication for conflicting cases.
In the reliability phase, the two annotators achieved Cohen's $\kappa$ values of $0.66$ (SATD identification), $0.63$ (subtype classification), and $0.62$ (event coupling classification), all indicating substantial agreement~\cite{landis_measurement_1977}.
This proves that the codebook is reliable, and the annotators can move on to label the rest of the dataset in the respective tasks.
Production phase shows similar levels of agreement as the reliability phase, and all the disagreements are manually resolved during a adjudication meeting between the two labellers.

\paragraph{Low statistical power for survival analysis.}
As shown in Table~\ref{tab:survival_logrank} of RQ2 (Section~\ref{sec:results-rq2}), survival analysis dataset contain only a few instances for several SATD subtypes, and limits the statistical power of log-rank tests.
This is in spite of our dataset construction via stratified prefix sampling which involves scrapping of Docker Hub by query the DockerHub API for all the projects that starts with all combination of 3-character prefixes.
However, this quantitative aspect of our study is only one half of the equation; we complement the quantitative analyses with the qualitative taxonomy study in RQ3 (Section~\ref{sec:results-rq3}), 
and presented concrete examples of source code-side co-evolution that trigger admissions or enable repayments of SATD in Dockerfiles.
This is in line with existing empirical studies in self-admitted technical debt repayments~\cite{muse_fixme_2022,zampetti_was_2018,bernardo_impact_2023} 
that strengthen limited quantitative time-to-event results with qualitative analyses.

\subsection{External validity.}
Our dataset comprises Docker Hub-GitHub linked open-source projects with at least one star, and our qualitative taxonomy is based on $411$ coupled SATD events from these projects.
Therefore, our qualitative findings are most relevant to these actively maintained open-source containerized projects.
We require extension to this work to validate our findings on closed-source images, other container ecosystems (e.g., Podman~\cite{noauthor_podman_nodate}, Buildah~\cite{noauthor_buildahio_nodate}), or to non-Docker IaC artifacts (e.g., Kubernetes manifests~\cite{penfound_production-grade_nodate}, Terraform configurations~\cite{noauthor_terraform_nodate}).
However, our work follows the already-established research topic of Dockerfiles~\cite{rosa_fixing_2024,bui_dockercleaner_2023,durieux_parfum_2023,rosa_not_2024,durieux_empirical_2024,wu_characterizing_2020}, and focuses on adding co-evolution perspective to provide more context to the study of SATD in Dockerfiles.

\section{Related Works}
\label{sec:related}

\subsection{SATD Identification and Classification}

Early approaches for SATD detection relied on keyword and pattern matching in source-code comments~\cite{guo_how_2021}.
Subsequent work uses NLP-based classifiers that could outperform such keyword-based heuristics for identifying design and requirement debt~\cite{maldonado_using_2017}.
Text-mining approaches with composite cross-project classifiers~\cite{huang_identifying_2018} and CNN-based models~\cite{ren_neural_2019} further improved detection accuracy and cross-project generalizability.
More recently works apply fine-tuned large language models for better identification performance~\cite{sheikhaei_empirical_2024}.
Finally, multi-task learning approaches leverage information from various artifacts such as code comments, issue trackers, pull requests, and commit messages~\cite{li_automatic_2023,gu_self-admitted_2024,li_impact_2025}, and jointly model SATD alongside other topics such as code vulnerabilities~\cite{russo_leveraging_2025}.
All of these detection and classification approaches have the limited scope of considering only the comment text and its immediately surrounding code.
\textbf{Our work builds on top of the keyword-based approach~\cite{guo_how_2021} for SATD candidate extraction, and adds a cross-artifact dimension that existing SATD detection approaches do not consider.}

\subsection{Production-Test Co-evolution}

Early works in production-test co-evolution established the notion that production and test code should ideally evolve synchronously~\cite{zaidman_mining_2008,zaidman_studying_2011,lubsen_using_2009}.
Subsequent studies mined fine-grained co-change patterns via association rule mining to identify source file type-specific relationships between production and test co-modification~\cite{marsavina_studying_2014,vidacs_co-evolution_2018}.
Levin and Yehudai~\cite{levin_co-evolution_2017} showed that even though developers frequently fix production code without updating the test cases, when test maintenance does occur, it correlates with semantic change in production code.
More recently, Wang et al.~\cite{wang_understanding_2021} built predictive models to detect outdated tests to identify when co-evolution is needed
while Sun et al.~\cite{sun_revisiting_2023} investigated the assumption underlying co-evolution sample construction to remove noise, and proposed a more reliable identification technique.
Lastly, approaches leveraging LLMs to automate co-evolution~\cite{chi_reaccept_2025,li_cocoevo_2025} use structural similarity metrics to generate co-evolved test cases~\cite{shimmi_leveraging_2022}.
\textbf{Our work applies this co-evolution premise from the production-test relationship to the Dockerfile-source code relationship, and investigates not only the prevalence, but also what reasons behind the co-evolutions.}

\subsection{Code-Comment Inconsistency}

Fluri et al.~\cite{fluri_analyzing_2009} empirically studied comment-code co-evolution across eight software systems, and found that even though code and comments co-evolve in roughly 90\% of cases, those changes are re-documented only in later revisions rather than in the commits where those co-evolution happens.
Such cases of deferred updates motivate the topic of code-comment inconsistency (CCI) that studies comments that do not faithfully describe their corresponding pieces of code.
Rong et al.~\cite{rong_code_2025} extended CCI research from detection towards automated rectification by fine-tuning a large language model to both identify and correct inconsistent comments. 
They reported that automated CCI resolution requires leveraging the semantic relationship between a comment and its surrounding code context.
\textbf{While CCI research study code comments within the files they are contained in, our study puts Dockerfile SATD into context of the commit-level view involving co-changed artifacts, shifting the paradigm from intra-file to cross-artifact.}

\subsection{IaC-Source Co-evolution}

Jiang and Adams~\cite{jiang_co-evolution_2015} showed that IaC specification files (Puppet/Chef) are large, churn frequently, and are tightly coupled with test files across OpenStack projects.
Wu et al.~\cite{wu_dockerfile_2020} characterized Dockerfile change patterns across 4,110 GitHub projects and reported how Dockerfiles co-change with other project files, although without examining the \emph{rationale} behind such co-changes.
These studies focus only on \emph{structural} co-change (whether files change together, and to what extend), and does not investigate the \emph{rationale} behind the co-changes. 
For example, they do not ask whether a change in an infrastructure file was triggered by a technical limitation acknowledged in the co-changed file
\textbf{Our work fills this rationale gap by using investigating co-change or co-evolution patterns as the rationale behind Dockerfile SATD admissions and repayments, thereby connecting the IaC co-evolution literature with the SATD lifecycle literature for the first time within our knowledge.}

\section{Conclusion and Future Work}
\label{sec:conclusion}

We studied self-admitted technical debt (SATD) in Dockerfiles through the lens of co-evolution.
We conducted a empirical study on a Docker Hub-GitHub dataset, by manually annotated SATD instances in Dockerfile and their coupling with other source files.
We used the annotated dataset to answer three research questions on the prevalence, coupling rates, survival rates, and characteristics source code-side changes that trigger or repay SATD in Dockerfiles.
We found that Dockerfile-only view is an incomplete unit of observation as approximately 27\% of admissions and 40\% of repayments events involve co-evolution with source code.
We also found that coupled SATD is repaid faster overall ($p=0.0201$) although \emph{Code/MissingFunctionality} coupled instances persist longer.
In future work, we plan to generalize this SATD-source code co-evolution paradigm to other IaC technologies such as Kubernetes and Terraform to investigate 
relationship between the prevalence of SATD and security vulnerabilities in IaC artifacts, and their automated remediations.
Moreover, researchers in existing software engineering topics such as automated SATD repayment, code smell repair, and code-comment inconsistency rectification 
can integrate the co-evolutionary perspective to overcome source code-related roadblocks discussed in this work to achieve breakthroughs in their respective areas.

\section*{Data Availability}
The dataset constructed and analyzed for this study is publicly availably online at: 
\url{https://osf.io/sh5xd/overview?view_only=e06572d75ee54348807f3925c14b0371}~\cite{minn_dockerfile_nodate}.

\bibliographystyle{spmpsci}      

\bibliography{references}   

\end{document}